\documentclass[12pt,preprint]{aastex}
\shorttitle{Classical Cepheid Pulsation Models}
\shortauthors{Bono et al.}

\begin{document}

\title{Classical Cepheid Pulsation Models. X. The Period-Age Relation}

\author{G. Bono\altaffilmark{1}, M. Marconi\altaffilmark{2}, 
S. Cassisi\altaffilmark{3}, F. Caputo\altaffilmark{1}, 
W. Gieren\altaffilmark{4}, G. Pietrzynski\altaffilmark{4,5}} 

\altaffiltext{1}{INAF - Osservatorio Astronomico di Roma, via Frascati 33,
00040 Monte Porzio Catone, Italy; bono@mporzio.astro.it, caputo@mporzio.astro.it}
\altaffiltext{2}{INAF - Osservatorio Astronomico di Capodimonte, via Moiariello 16, 80131 Napoli, Italy; marcella@na.astro.it}
\altaffiltext{3}{INAF - Osservatorio Astronomico di Teramo, via M. Maggini, 64100 
Teramo, Italy; cassisi@te.astro.it} 
\altaffiltext{4}{Universidad de Concepcion, Departamento de Fisica, Astronomy 
Group, Casilla 160-C, Concepcion, Chile; wgieren@coma.cfm.udec.cl, 
pietrzyn@hubble.cfm.udec.cl} 
\altaffiltext{5}{Warsaw University Observatory, Al. Ujazdowskie 4, 00-478 
Warszawa, Poland} 

%%%%%%%%%%%%%%%%%%%%%%%%%%%%%%%%%%%%%%%%%%%%%%%%%%%%%%%%%%%%%%%%%%%%%%%%%%%

\pagebreak 
\begin{abstract}
We present new Period-Age (PA) and Period-Age-Color (PAC) relations for 
fundamental and first overtone classical Cepheids. Current predictions 
rely on homogeneous sets of evolutionary and pulsation models covering  
a broad range of stellar masses and chemical compositions. We found that 
PA and PAC relations present a mild dependence upon metal content. Moreover,
the use of different PA and PAC relation for fundamental and first overtone 
Cepheids improves the accuracy of age estimates in the short-period 
($\log P < 1$) range (old Cepheids), because they present smaller intrinsic 
dispersions. At the same time, the use of the PAC relations improves the 
accuracy in the long-period ($\log P \ge 1$) range  (young Cepheids), 
since they account for the position of individual objects inside the 
instability strip. 

We performed a detailed comparison between evolutionary and pulsation 
ages for a sizable sample of LMC (15) and SMC (12) clusters which host 
at least two Cepheids. In order to avoid deceptive uncertainties in the 
photometric absolute zero-point, we adopted the homogeneous set of 
B,V,I data for clusters and Cepheids collected by OGLE. We also 
adopted the same reddening scale. The different age estimates 
agree at the level of 20\% for LMC clusters and of 10\% for 
SMC clusters. We also performed the same comparison for two 
Galactic clusters (NGC6067, NGC7790) and the difference in age 
is smaller than 20\%.  

These findings support the use of PA and PAC relations to supply  
accurate estimates of individual stellar ages in the Galaxy and 
in external Galaxies. The main advantage of this approach is 
its independence from the distance.     
\end{abstract} 


%%%%%%%%%%%%%%%%%%%%%%%%%%%%%%%%%%%%%%%%%%%%%%%%%%%%%%%%%%%%%%%%%%%%%%%%%%%%%
%\pagebreak  
\section{Introduction}

Classical Cepheids are crucial objects in several long-standing astrophysical 
problems. These variable stars, due to their intrinsic luminosity and regular 
luminosity variation, can easily be detected in the outskirt of the Galactic 
disk and in external galaxies. Cepheids are the most popular primary 
distance indicators, since the pulsation periods and the colors are connected 
with the luminosity, and therefore the Period-Luminosity (PL) and the 
Period-Luminosity-Color (PLC) relation can be adopted for estimating cosmic 
distances (Feast 1999; Udalski et al. 1999a,b,c; Freedman et al. 2001; 
Saha et al. 2001; Fouqu\'e et al. 2003; Tammann, Sandage, \& Reindl 2003;
Storm et al. 2004). 

Classical Cepheids cover the whole range of intermediate-mass stars, and 
thus it is not surprising that these objects are typically adopted to 
trace Population I stars in the Galactic disk (Kraft \& Schmidt 1963), 
and star-forming regions in extragalactic systems (Elmegreen \& Efremov 1996). 
The use of Cepheids as tracers of young stellar populations was supplemented 
by the evidence that, if these objects obey to a PL relation,  they also 
obey to a Period-Age (PA) relation. Therefore, by taking into account the 
Mass-Luminosity relation and the stellar ages predicted by evolutionary 
models, it was soon recognized (Kippenhahn \& Smith 1969; Meyer-Hofmeister 1969) 
that an increase in the pulsation period implies an increase in the stellar mass, 
and in turn a decrease in the Cepheid age. On the basis of these plane physical 
arguments Efremov (1978) derived an empirical PA relation by adopting 
Cepheids in Galactic, in M31, and in Large Magellanic Cloud (LMC) clusters,  
whose age was independently estimated (see also Tsvetkov 1989). More 
recently, Magnier et al. (1997, hereinafter M97) derived a new semi-empirical 
PA relation and used Cepheids in NGC206, the super-association in M31, 
to trace the age distribution, and in turn the star formation history  
in this region, located at the intersection of two spiral arms. A similar 
approach was also adopted by Efremov \& Elmegreen (1998, hereinafter EE98) 
and by Grebel \& Brandner (1998, hereinafter GB98), who also derived new 
PA relations on the basis of a larger sample of cluster Cepheids. 
By relying on these new relations, and on the large photometric database 
current available for LMC Cepheids -more than 1000 objects- they provided 
useful constraints on the star formation history of LMC over the last 
200 Myr. 

The main advantages in using the PA relation for estimating the 
Cepheid ages are the following:
{\em i)} the age estimates rely upon an observable -the period- which is 
marginally affected by systematic errors such as reddening, distance, 
and photometric calibrations. {\em ii)} The PA relation can be applied 
to individual objects. Therefore, relative age estimates, based on this 
method, can supply accurate constraints on the occurrence of age 
gradients, since field Cepheids outnumber open clusters.   
{\em iii)} The application of 
the PA relation to cluster Cepheids supply the unique opportunity to 
estimate the age of the parent cluster, even though the photometry of 
main sequence turn-off stars is lacking or uncertain.  
However, empirical PA relations are affected by three drawbacks:
{\em i)} current PA relations have been calibrated by using cluster 
Cepheids. However, both the zero-point and the slope of these relations 
are affected by uncertainties in the distance modulus and in the 
reddening of selected clusters, as well as by the period range covered 
by cluster Cepheids. {\em ii)} Cepheid PA relations have only been 
derived for fundamental Cepheids. The possibility to derive independent 
PA relations for fundamental and first overtone Cepheids would improve 
the accuracy of the PA relation in the short-period ($\log P \le 1$) 
range, because the 
intrinsic scatter is smaller. {\em iii)} Cepheid ages based on the 
PA relations rely on the statistical  assumption that the Cepheid 
instability strip has a negligible temperature width. This is a 
plausible assumption in the short-period range, but the ages of 
long-period ($\log P > 1$) Cepheids should be estimated on the basis of a   
Period-Age-Color (PAC) relation. This relation accounts for individual 
Cepheid positions inside the instability strip, and therefore it is 
also marginally affected by the spread in luminosity due to 
evolutionary effects (M97).   
The reader interested in a detailed discussion concerning the width 
in temperature of the instability strip is referred to Bono et al. (2000a) 
and to Bono, Caputo, \& Marconi (2001).

The main aim of this investigation is to supply a new theoretical 
calibration of both PA and PAC relations for fundamental and first 
overtone Cepheids, by adopting homogeneous and detailed sets of 
nonlinear, convective pulsation models together with up-to-date 
evolutionary models. In order to constrain the accuracy of the new 
relations, we also plan to perform a detailed comparison between 
Cepheid pulsation ages and ages based on the fit of isochrones to 
turn-off stars of the parent cluster. This test will be performed 
on a sizable sample of Magellanic clusters hosting more than two 
Cepheids and two Galactic clusters. 

In section 2, we describe the evolutionary and pulsational frameworks 
adopted in this investigation together with the procedures for computing 
current PA and PAC relations. In section 3, we present photometric data for 
cluster and Cepheid stars, and discuss the comparison between cluster and 
Cepheid ages. Finally, in section 4, we summarize our results and briefly 
outline future developments.

%%%%%%%%%%%%%%%%%%%%%%%%%%%%%%%%%%%%%%%%%%%%%%%%%%%%%%%%%%%%%%%%%%%%%%%%%%%%
\section{Theoretical models} \label{phys}

\subsection{Stellar evolutionary models}

The current evolutionary framework is based on the updated set of stellar 
evolutionary models recently computed by Pietrinferni et al. (2004). 
The reader interested in a detailed discussion concerning the input 
physics is referred to this paper. The main changes when compared 
with our previous investigation on intermediate-mass evolutionary 
models (Bono et al. 2000b) are the following:
 
\begin{itemize} 

\item {\em Opacity - }  
Conductive opacity for electron degenerate matter was estimated following 
the prescriptions by Potekhin (1999).
 
\item {\em Energy loss rates - }  
The energy loss rates for plasma-neutrino processes have been updated by
using the most recent and accurate estimates provided by Haft, Raffelt
\& Weiss~(1994).  For the other processes, we follow the prescriptions 
adopted by Cassisi \& Salaris (1997).
 
\item {\em Nuclear reaction rates - }  
The nuclear reaction rates have been updated by using the NACRE database 
(Angulo et al. 1999), with the exception of the 
$^{12}$C$(\alpha,\gamma)^{16}$O reaction. For this reaction we adopt the 
recent determination by Kunz et al. (2002).  Electron screening is 
treated according to Graboske et al. (1973);
 
\item {\em EOS - }  
The updated Equation of State (EOS) by A. Irwin\footnote{The EOS code
has been made available at ftp://astroftp.phys.uvic.ca under the 
GNU General Public License (GPL)} has been used. It is noteworthy
that this EOS allowed us to compute self-consistent stellar models in 
all evolutionary phases relevant for this investigation. For a thorough
discussion concerning the key features of this EOS, see e.g. Cassisi, 
Salaris, \& Irwin (2003) and Irwin et al. (2004, in preparation). 

\item {\em Convection  - }  
Superadiabatic convection is handled according to the Cox \& Giuli (1968) 
formalism of the mixing length theory (B\"ohm-Vitense 1958). The mixing 
length parameter has been fixed according to the solar calibration 
($ml=1.713$). \\  
The efficiency of the physical mechanisms that govern the size of the 
convective core during central H-burning phases is still an open problem  
(see, e.g., Testa et al. 1999; Maeder \& Meynet 2000; Barmina, Girardi, 
\& Chiosi 2002; Brocato et al. 2003). Current stellar models do not 
account for core overshooting.  The reader interested in a detailed 
discussion concerning the impact of mixing processes in intermediate-mass 
stars is referred to Castellani et al. (1985), Chiosi \& Maeder (1986), 
and Cassisi (2004).
 
\item {\em Chemical composition - }  
The sets of evolutionary models have been computed for Z=0.004, 0.008, 
and Z=0.0198 (solar). The initial He abundance was chosen according 
to  Salaris et al. (2004) equal to $Y=0.245$. To reproduce the 
calibrated initial solar He-abundance we adopted $dY/dZ\approx1.4$ 
(Pietrinferni et al. 2004). The adopted He abundances for the three 
selected metal abundances are: Y=0.251, 0.256, and 0.273.

\item {\em Stellar mass - }  
The mass values of the evolutionary models adopted in this investigation 
range from $2.3M_\odot$ to $10M_\odot$. The mass loss is treated according 
to the Reimers formula (Reimers 1975) with $\eta$ equal 
to 0.4\footnote{The entire evolutionary database can be found at the 
following URL address: http://www.te.astro.it/BASTI/index.php.}.
\end{itemize} 

Fig. 1 shows the H-R diagram for stellar models adopted in this 
investigation. Stellar isochrones have been computed with steps of 
50 Myr for ages older than 200 Myr and of 20 Myr for younger ages.
Theoretical predictions have been transformed into the observational 
plane by adopting the bolometric corrections and the color-temperature 
transformations provided by Pietrinferni et al. (2004).  
The current set of stellar isochrones covers the typical age range 
of intermediate-mass stars with a very good resolution. Therefore,  
we also estimated the dependence of age on intrinsic parameters. 
In particular, we found: 

\[
\log t = (8.911\pm 0.018) + (0.423\pm0.002)\times M_V^{TO} + 
(0.200\pm0.008)\times log Z \;\;\;\;\; r^2=0.99 \;\;\;  [1]  
\]

where t is the age in yr, $M_V$ the TO visual magnitude, $r$ the correlation 
parameter, and the other symbols have their usual meaning. Note that current 
analytical relation applies to stellar structures with ages ranging from 
30 Myr to 1 Gyr, and for the selected chemical compositions (see Table 5).

%%%%%%%%%%%%%%%%%%%%%%%%%%%%%%%%%%%%%%%%%%%%%%%%%%%%%%%%%%%%%%%%%%%%%%%%%%%%
\subsection{Pulsation models} \label{phys}

We adopted the sequences of nonlinear, convective pulsation models constructed 
by Bono et al. (1999, 2000a,2002a) to estimate the PA and the PAC relations. 
This set of models was computed by adopting chemical 
compositions typical of Galactic (Y=0.28, Z=0.02) and Magellanic Cepheids 
(Y=0.25, Z=0.004, Small Magellanic Cloud [SMC]; Y=0.25, Z=0.008, Large 
Magellanic Cloud [LMC]) and cover a wide range of stellar masses and 
effective temperatures. Note that the He abundances adopted to construct 
the pulsation models present a mild difference when compared with the 
evolutionary ones (see \S 2.1). However, such a difference has a marginal 
impact upon pulsation properties, and in turn on the conclusions of our 
current investigation. The input parameters adopted for fundamental 
and first overtone models are listed in Table 1. Both canonical and 
noncanonical Cepheid models were constructed by adopting the same 
mass-luminosity (M/L) relations used by Bono et al.  (1999). 
The grid of canonical Cepheid models was implemented with new 
sequences for $M/M_\odot=9.4, \log L/L_\odot=4.4$, $Z=0.004$ and 
for $M/M_\odot=10.1,  \log L/L_\odot=4.4$ and $Z=0.008$. 
On the basis of these models, we performed 
a linear least-squares fit to luminosities and effective temperatures, 
i.e. $\log T_e = \alpha + \beta \log L/L_\odot$, of blue and red edges 
of both fundamental and first overtone instability strip. The analytical 
relations for the three different chemical compositions are listed in 
Table 2. Note that the periods of fundamental pulsators range 
from $\log P = 0.17$ to $\log P = 2.02$, while for first overtones they 
range from $\log P = -0.20$ to $\log P = 0.88$. This Table also gives 
the analytical relations for the effective 
temperature of the fundamental instability strip at solar chemical 
composition provided by Petroni et al. (2003). These models have been 
constructed by adopting the same treatment for turbulent convection, but 
with different assumptions concerning input physics and spatial resolution 
across the ionization regions. For these
models a M/L relation based on more recent canonical evolutionary models
was also adopted. A glance at the two different sets of predictions  
(see Table 2) indicates that the difference is at most of the order of 
0.02 dex in the zero-point and smaller than 0.01 dex in the slope. This 
finding suggests that the temperature of the edges marginally depends 
on these parameters.  
However, recent empirical evidence (Tammann, Sandage, \& Reindl 2003;
Kanbur \& Ngeow 2004; Sandage, Tammann, \& Reindl 2004) suggests that the
PL relations for fundamental Cepehids, and in turn, the edges of the Cepheid
instability strip show a change in the slope at $\log P\approx 1$. According
to theoretical predictions based on convective models, this occurrence is
caused by a twofold effect. In the long-period range ($\log P > 1.5$) the
instability strip moves toward redder colors causing a flattening in the
PL relation (Bono \& Marconi 1998; Bono et al. 2000a). This behavior is
supported by observations (Sasselov et al. 1997). In the short-period
range ($\log P < 0.4$), the slope of the fundamental red edge becomes
shallower, and in turn, the width in temperature of the instability strip
becomes narrower (see Fig. 2 in Bono et al. 2001). This trend is also
supported by observations (Bauer et al. 1999; Udalski et al. 1999a,b).
To account for these effects, we also computed quadratic analytical relations
connecting the effective temperature of the instability edges to the stellar
luminosity (see Table 3). The quadratic fit has only been performed for
canonical models, since for this set we computed a finer grid of models.
Note that the difference between linear and quadratic relations is at most
0.05 dex, and therefore the impact on the PA and the PAC relations is negligible.
Obviously, the difference increases as soon as we move from the theoretical to
the observational plane, due to changes in the bolometric correction. We did not
estimate quadratic relations for first overtones, because both theory and
observations suggest that the instability edges for these pulsators are linear.
The difference between fundamental and first overtone Cepheids is due to
the narrower period and color range covered by the latter pulsators. 

The current analytical relations and evolutionary tracks (see \S 2.1) 
were adopted to select the evolutionary phases falling  
inside the instability strip. This means that we selected not only 
the second and the third crossing, but also the first one. 
Finally, the PA relations were estimated as a linear fit by weighing 
each individual evolutionary phase for the amount of time spent 
inside the instability strip. The coefficients of the PA relations 
as a function of the chemical composition are listed in Table 4. 
Fig. 2 shows the new PA relations for fundamental (bottom) and 
first overtone (top) Cepheid models. Data plotted in the bottom 
panel disclose that the dependence on the metal content is strictly 
decreasing when  moving from short (old) to long-period (young) 
Cepheids. On the other hand, data for first overtone pulsators 
(top panel) display a different behavior for periods longer than 
$\log P \ge 0.5$. This change is due to the narrowing in temperature 
of the first overtone instability strip when moving toward higher 
luminosities (Bono et al. 2002b). 
This prediction agrees quite well with observations, and indeed 
the period distribution of the first overtone pulsators in the Magellanic 
Clouds shows a well-defined cut-off toward longer periods 
(Udalski et al. 1999a,b; Beaulieu \& Marquette 2000). This region of 
the instability strip is a robust observable, since it can be adopted 
to constrain the luminosity above which Cepheids only pulsate in the 
fundamental mode (Bono et al. 2002b).  
Moreover, the metallicity effect among the different first overtone 
PA relations appears negligible, within the intrinsic dispersions. 

Fig. 3 shows the comparison between current predictions for  
solar chemical composition and similar relations available 
in the literature. The PA relations plotted in this figure 
show a reasonable agreement, in particular if we account for the 
fact that they have been derived using different approaches to 
calibrate the absolute zero-point, as well as different sets of 
evolutionary models. As expected, our PA relation supplies, at 
fixed period, younger ages because the evolutionary tracks we 
adopted do not account for convective overshooting. \\  
To estimate on a quantitative basis the age difference between 
evolutionary tracks which either account for or neglect convective 
core overshooting, the current set of stellar models has been implemented 
with an additional set of evolutionary tracks with stellar masses 
ranging from 4 to 10 $M_\odot$. These tracks were constructed by adopting 
a significant amount of core overshooting ($\lambda_c=0.2H_p$\footnote{The 
extent of the non-canonical mixed region is usually defined in terms of 
a parameter, $\lambda$, which fixes the length, in units of the local pressure 
scale height $H_p$, covered by convective cells into the stable region surrounding 
the convective core.}) and $\lambda_e$ (envelope 
overshooting\footnote{It is worth noting that envelope overshooting at odds with  
core overshooting, has no effect on the evolutionary lifetimes, but 
it substantially affects the morphology of the blue loops during the core 
He-burning phases.}) ranging from 0 to 0.5 $H_p$ during hydrogen and helium 
burning phases (Pietrinferni et al. 2004).    
For each given mass value and chemical composition, we estimated 
a mean luminosity and a mean temperature typical of the instability 
strip, i.e. $3.81 \le \log T_e \le 3.85$, according to the shape of 
the blue loop. On the basis of the predicted stellar mass, luminosity, and 
effective temperature, we estimated a mean fundamental period for both 
canonical and noncanonical models using the pulsation relations 
provided by Bono et al. (2000a). Data plotted in Fig. 4 (see also 
Table 5) indicate that the difference in age for the selected stellar 
masses ranges from 7\% ($8 M_\odot$, Z=0.008, Y=0.25) to 18\% 
($10 M_\odot$, Z=0.008, Y=0.25). Note that for $M= 4 M_\odot$, Z=0.004 
and $M= 5 M_\odot$, Z=0.02 the noncanonical evolutionary tracks do not 
perform the loop. Interestingly enough, 
these are the typical differences between current PA relations and those 
based on evolutionary models which account for convective core overshooting 
(M97; GB98). 

To further improve the accuracy of Cepheid age estimates, in 
particular in the long-period range, we computed new analytical 
relations accounting for the width in temperature of the 
instability strip. The coefficients of the PAC relations are 
listed in Table 6 as a function of the chemical composition. 
Note that current PAC relations have been computed, for each 
given chemical composition, for optical colors, namely B-V and 
V-I. Fig. 5 shows the projection onto a plane of the new 
theoretical  PAC relations. 
To estimate the difference between Cepheid ages based on the PA and on the
PAC relation we computed the age difference for each given composition along
two isoperiodic lines at $\log P=1$ and $\log P=1.5$. Table 7 gives for
the two selected periods the age based on the PA relation (columns 2,5,8),
and the age based on the PAC relation along the blue (columns 3,6,9) and
the red (columns 4,7,10) edge of the instability strip. Table 7 also gives
the logarithmic temperatures (number in parentheses) of the instability edges
in which the period is equal to $\log P=1$ or 1.5. Data listed in this table
indicate that the age difference ranges from $\Delta \log t=0.04 - 0.07$
for hotter Cepheids to $\Delta \log t = -0.01 - -0.04$ for cooler ones.

%%%%%%%%%%%%%%%%%%%%%%%%%%%%%%%%%%%%%%%%%%%%%%%%%%%%%%%%%%%%%%%%%%%%%%%%%%%%%%
\section{Empirical data} 
In order to validate current PA and PAC relations, we decided to perform 
a detailed comparison between age estimates based on the canonical 
Main-Sequence fitting, and on periods and colors of cluster Cepheids. 
Photometric data collected by OGLE for Magellanic cluster 
Cepheids (Udalski et al. 1999a,b; Pietrzynski \& Udalski 1999) are very 
useful to accomplish this goal. These data are very homogeneous, since 
they have been collected with the same observational equipment, and  
reduced by adopting the same approach. Moreover, and even more importantly, 
static and variable stars have the same absolute photometric zero-point. 
To perform the test we selected LMC and SMC clusters hosting at least 
two cluster Cepheids. We ended up with a sample of 15 and 12 clusters  
respectively.   

Figures 6a,d and 7a,c\footnote{The figures 6b,c,d and 7b,c are only available 
in the on-line edition of the manuscript.} show the fit of stellar 
isochrones at fixed metal-content with cluster stars in two different 
Color-Magnitude Diagrams, namely V,B-V and V,V-I. Circles and squares  
mark fundamental and first overtone cluster Cepheids, according to the 
classification and the membership provided by Pietrzynski \& 
Udalski (1999). Individual cluster age estimates have been performed by 
assuming the same distance modulus (18.5 for LMC clusters and 19.00 for 
SMC clusters), and the mean 
reddening estimate based on cluster Cepheids provided by Pietrzynski \& 
Udalski (1999). It is worth mentioning that we did not force stellar 
isochrones to fit both magnitude and colors of classical Cepheids. 
The extent in color of the so-called blue loop depends on several 
physical parameters and assumptions adopted to construct intermediate-mass 
evolutionary models (Chiosi, Bertelli, \& Bressan 1992; Stothers \& 
Chin 1993,1994; Bono et al. 2000b). As a whole, the agreement between the 
fit in the  V,B-V  and in the V,V-I planes is quite good. The color 
excess in V-I was estimated from E(B-V) according to Cardelli et al. 
(1988). Table 8 lists the new cluster age estimates, together with 
the mean reddening value. To provide a plausible estimate of the formal 
error on isochrone ages, we assumed an uncertainty of 0.15 mag on the 
distance that accounts for uncertainty on the individual cluster distances 
and on reddening estimates and of 0.2 dex on the cluster metallicity. 
By accounting for these uncertainties and the intrinsic errors on the 
coefficients of equation 1, we end up with a tipical uncertainty on the 
cluster age of the order of 0.075 dex.  

At the same time, we estimated the cluster ages using both the PA 
and the PAC relations for fundamental and first overtone Cepheids. 
Individual cluster ages are listed in Table 8, together with the 
formal uncertainties given by the uncertainties on the coefficients 
of PA and PAC relations. Figures 8 and 9 display the relative age 
difference between evolutionary (isochrones) and pulsation ages 
for LMC and SMC clusters, respectively. Interestingly enough, data 
plotted in these figures show that, within current empirical and 
theoretical uncertainties (see error bars), evolutionary and 
pulsation cluster ages agree on average at the level of 20\% for 
LMC clusters and 10\% for SMC clusters.
The difference between the two sets, and in particular the mild increase in
the LMC Cepheid ages based on the PAC relation (top panel Fig. 8), might be due
to uncertainties in the reddening correction (Zaritsky et al. 2002, and references
therein) and/or in the cluster membership, since LMC clusters are typically located
across the bar. 
   
In order to provide a comprehensive analysis over a broad metallicity range,  
we also investigated two Galactic clusters hosting at least two 
classical Cepheids, namely NGC~6067 and NGC~7790. By adopting periods, 
mean magnitudes, and individual reddenings provided by Berdnikov (2000) 
for Cepheids in NGC~7790 (CEa Cas, CEb Cas, CF Cas) and by 
Laney \& Stobie (1994) for Cepheids in NGC~6067 (QZ Nor, V430 Nor) 
we found that mean cluster ages are $\log t = 7.85\pm 0.02$ yr (PA) and  
$7.81\pm 0.02$ yr (PAC) for NGC~7790, as well as $\log t = 7.76\pm 0.22$ yr (PA) 
and  $7.79\pm 0.11$ yr (PAC) for NGC~6067 (see Table 9). It is worth noting 
that Romeo et al. (1989), by using a period-age relation based on evolutionary 
tracks accounting for convective core overshooting and mass loss, estimated 
an age of $\log t = 8.1$ yr for Cepheids in NGC~7790.  

As far as the cluster photometry is concerned, we adopted  the B,V,I data 
available in the web site developed and maintained by J.-C. Mermilliod 
for stellar systems \\ (http://obswww.unige.ch/webda/webda.html). Original 
B,V data for NGC~6067 have been collected by Walker \& Coulson (1985), 
while V,I data have been collected by Piatti, Claria, \& Bica (1998). 
At the same time, B,V,I data NGC~7790, have been collected by Phelps, \& 
Janes (1994), Gupta et al. (2000), and Henden (2003, \\  
ftp://ftp.nofs.navy.mil/pub/outgoing/aah/sequence/). Fig. 10 shows the 
fit between empirical data and stellar isochrones at solar chemical composition. 
The cluster distances adopted to perform the fit were estimated using the 
recent calibration of the optical PL relations for Galactic Cepheids, recently 
provided by Storm et al (2004), and the apparent magnitudes and reddenings listed 
in Table 9. We found that the true distance moduli are 
$11.27\pm 0.35$ for NGC~6067 and $12.50\pm0.23$ for NGC~7790. By adopting 
these distances and mean cluster reddenings of 0.32 (NGC~6067, Coulson \& 
Caldwell 1985) and 0.53 (NGC~7790, Lee \& Lee 1999), we estimated cluster 
ages of $70\pm10$ and $80\pm 10$ Myr, respectively. Once again, the 
difference between pulsational and evolutionary ages is smaller than 20\%. 

%%%%%%%%%%%%%%%%%%%%%%%%%%%%%%%%%%%%%%%%%%%%%%%%%%%%%%%%%%%%%%%%%%%%%%%%%%%%

\section{Summary and final remarks} 

We computed new PA and PAC relations for fundamental and first overtone 
Cepheids for chemical compositions typical of Magellanic and Galactic 
Cepheids. The current predictions rely on homogeneous sets of stellar isochrones 
and pulsation models covering the entire range of intermediate-mass stars.  
We found that the metallicity has a marginal effect on the PA relation. 
To validate current theoretical scenario, we performed a detailed comparison 
between evolutionary (isochrone fit) and pulsation (PA and PAC relations) 
ages of a sizable sample of LMC and SMC clusters hosting at least two 
Cepheids. In order to avoid deceptive uncertainties in the photometric calibration 
of both Cepheids and cluster stars, we adopted the homogeneous B,V,I photometric 
catalog provided by the OGLE collaboration (Pietrzynski \& Udalski 1999). 
The isochrone fit was performed both in the V,B-V and in the V,V-I CMDs
by adopting the reddening estimated by Pietrzynski \& Udalski (1999) for 
cluster Cepheids. We found that pulsational and evolutionary cluster 
ages agree quite well, and indeed the difference in age ranges from 10\% 
to 20\%.  We performed the same comparison for two Galactic clusters,  
namely NGC6067 and NGC7079. We found once again that the difference 
between pulsation and evolutionary ages is smaller than 20\%.  

These findings support the evidence that PA and PAC relations supply 
accurate individual age estimates. This means that they can be adopted 
to trace the age distribution not only in the Galactic disk, but also 
across the main body of the Magellanic Clouds. It is worth mentioning
that the use of independent relations for fundamental and first overtone 
Cepheids will supply more accurate age estimates in the short-period 
($\log P < 1$) range, since the intrinsic dispersion is smaller. 
At the same time, the use of the PAC relations will supply more 
accurate age estimates in the long-period ($\log P > 1$) range. 
Note that the vertical error bars in Fig. 8 and 9 mainly depend on the
number of cluster Cepheids. Individual Cepheid ages based on the PAC relation
are more accurate than the ages based on the PA relation, 
as soon as the uncertainty on the reddening correction is on average smaller
than 0.08 mag for B-V colors and 0.07 mag for V-I colors. These limits have
been estimated by accounting for the color coefficients listed in Table 6, 
and the age differences between PA and PAC relations listed in Table 7. 
The theoretical framework we developed can be adopted to detect the 
occurrence of age gradient(s), not only across star-forming regions, but 
also along the spiral arms of external galaxies. The sizable sample of 
extragalactic Cepheids detected by the two HST Key Projects aimed at 
estimating the Hubble constant (Freedman et al. 2001; Saha et al. 2001)
appears a well-suited extension. The same applies to the large sample 
of Cepheids discovered in M31 and M33 by the DIRECT experiment 
(Bonanos et al. 2003, and references therein).       

Finally, we would also like to mention that well-observed open clusters 
hosting sizable Cepheid samples can be safely adopted to test the accuracy 
of the physical ingredients currently adopted to construct evolutionary 
models. The difference between the cluster age based on the luminosity 
function and the pulsation ages based on PA and PAC relations can supply 
tight constraints on He-burning phases. Note that these two methods do 
not depend at all upon the cluster distance.  

\acknowledgments 
It is a pleasure to thank V. Castellani for several thorough discussions 
and a detailed reading of an early draft of this manuscript. We are also 
very grateful to A. Bragaglia for many useful suggestions concerning the 
photometric data of Galactic open clusters. We wish also to thank an 
anonymous referee for his/her detailed suggestions and helpful comments 
which improved both the content and the readability of this paper. 
We acknowledge J.-C. Mermillod and A. Henden for making available the 
photometric data bases they collected. 
This work was partially supported by MIUR/COFIN~2002 under the project 
(\#028935): "Stellar Populations in Local Group Galaxies" and by INAF 
under the project: "The Large Magellanic Cloud". WG and GP gratefully 
acknowledge financial support for this work from the Chilean Center 
for Astrophysics FONDAP 15010003.   

%%%%%%%%%%%%%%%%%%%%%%%%%%%%%%%%%%%%%%%%%%%%%%%%%%%%%%%%%%%%%%%%%%%%%%%%%

%%%%%%%%%%%%%%%%%%%%%%%%%%%%%%%%%%%%%%%%%%%%%%%%%%%%%%%%%%%%%%%%%%%%%%%%%%%%%
% Fig. 1 
\clearpage
\begin{figure}
%\epsscale{0.80}
\plotone{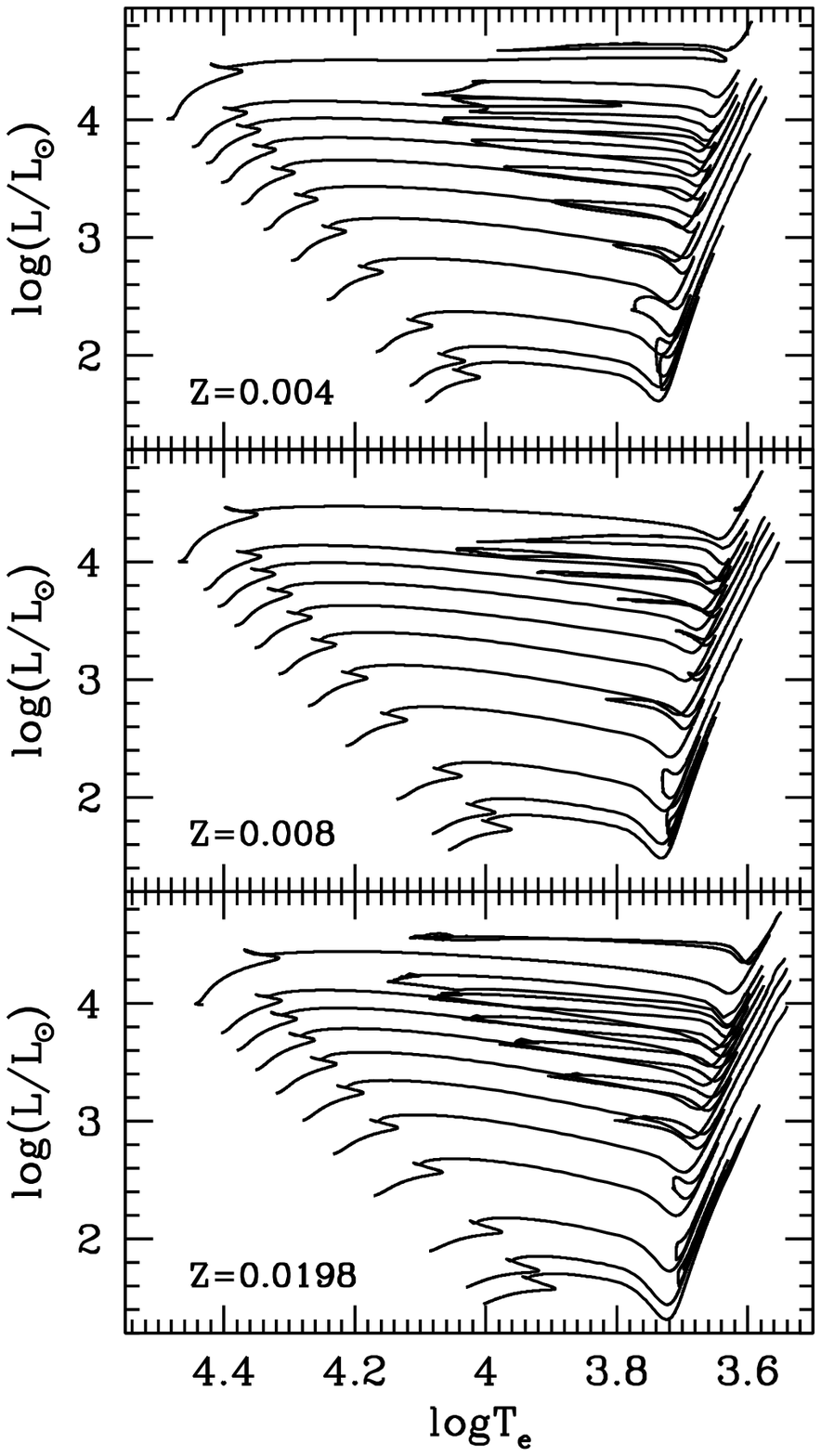}
\caption{Theoretical H-R diagram for intermediate-mass stars 
with different chemical compositions (see labeled values). See 
text for more details.\label{fk}}
\end{figure}

% Fig. 2 
\clearpage
\begin{figure}
%\epsscale{0.80}
\plotone{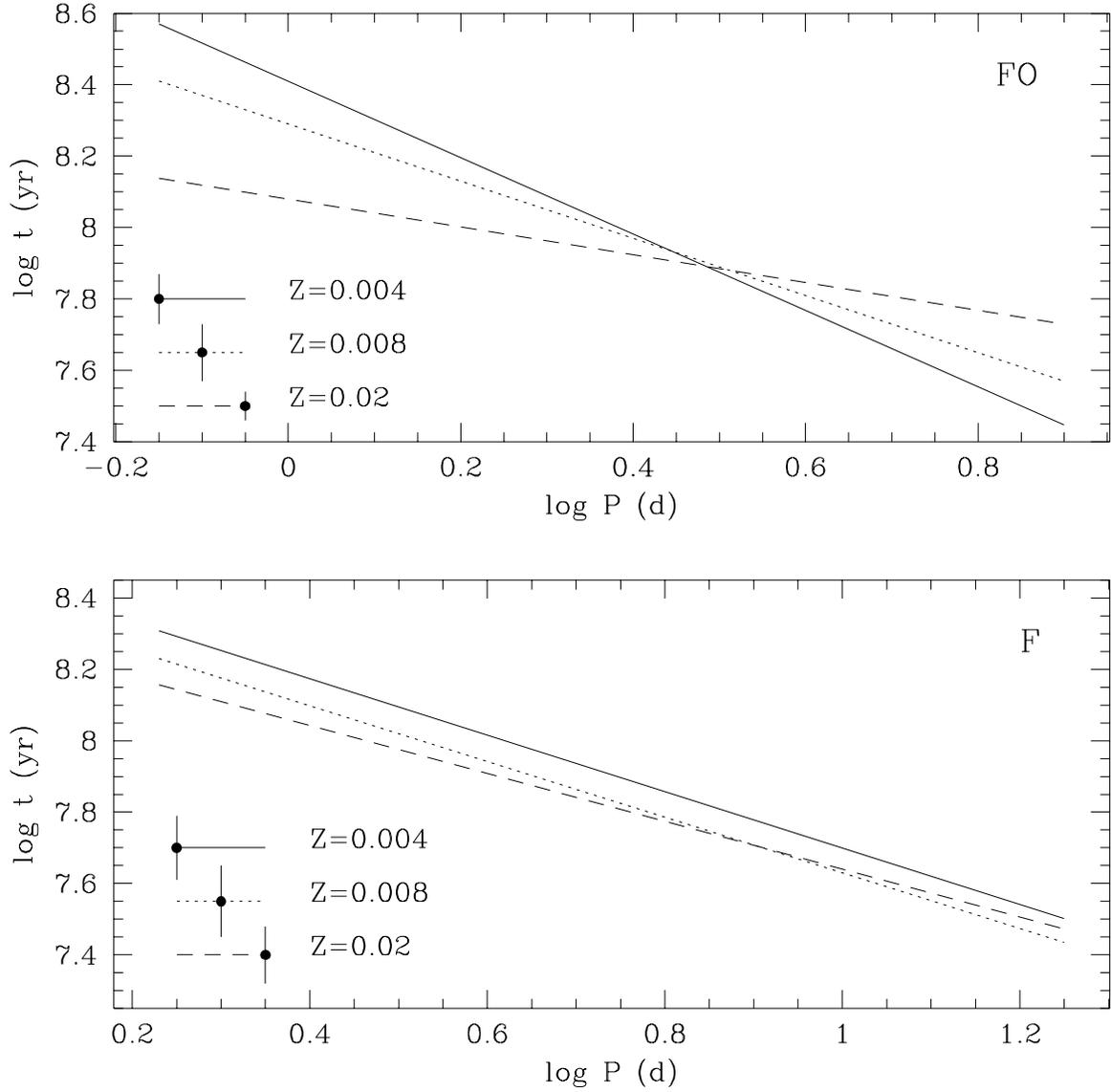}
\caption{Predicted Period-Age relations for first-overtone (top) and 
fundamental (bottom) Cepheids with different chemical compositions (see 
labeled values). The vertical bars display the standard deviation 
of analytical relations.\label{fk}}
\end{figure}

% Fig. 3 
\clearpage 
\begin{figure}
%\epsscale{0.80}
\plotone{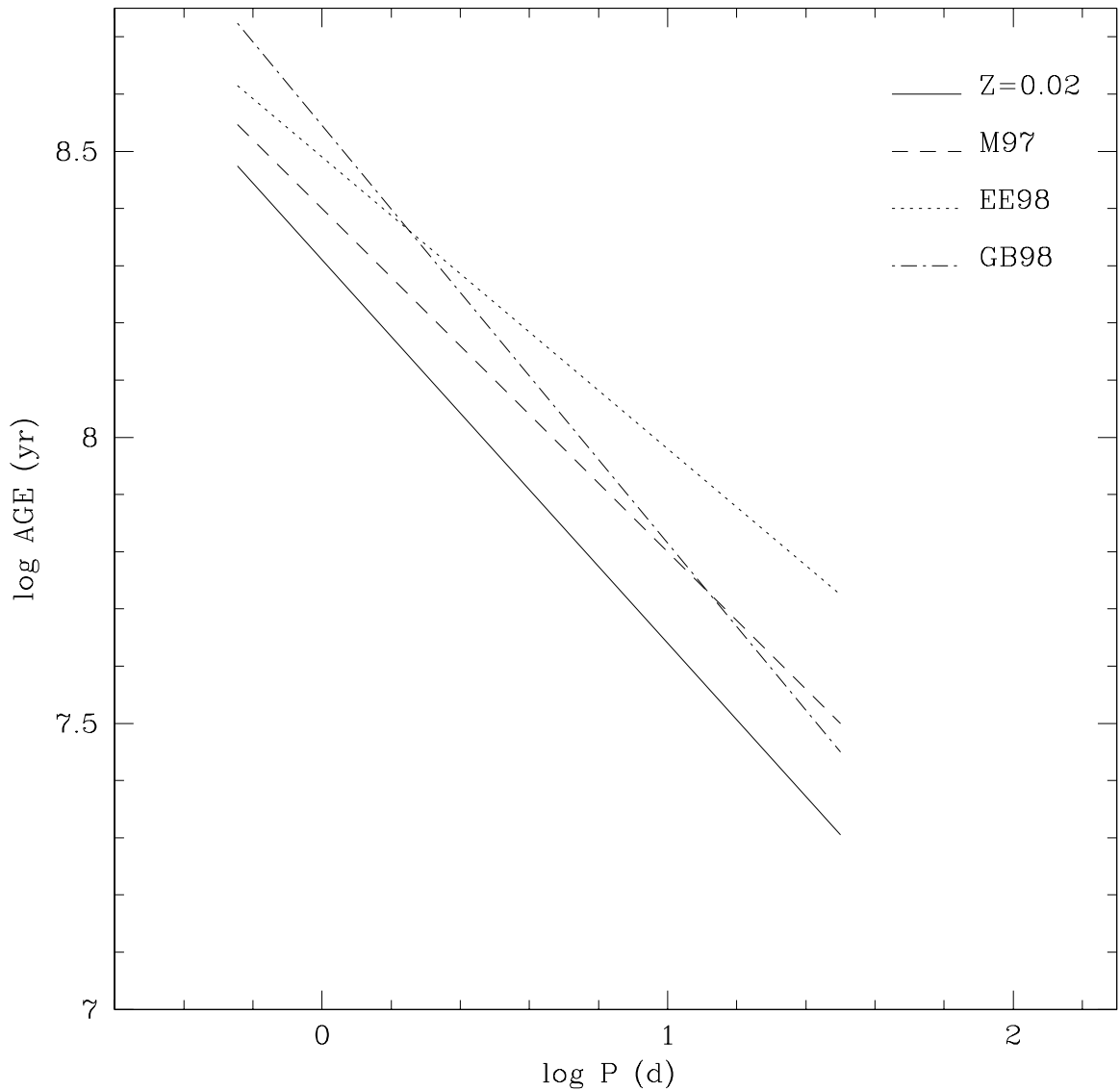}
\caption{Comparison between current Period-Age relation at solar chemical 
composition (solid line) and similar relations available in the literature. 
The dashed line shows the PA relation provided by Magnier et al. (1997), 
while the dotted and the dashed-dotted lines the PA relations by Efremov \&
Elmegreen (1998) and by Grebel \& Brandner (1998), respectively. Note that 
PA relations plotted in this figure have been obtained using different 
evolutionary models. See text for more details.\label{feosI}}
\end{figure}

% Fig. 4 
\clearpage 
\begin{figure}
%\epsscale{0.80}
\plotone{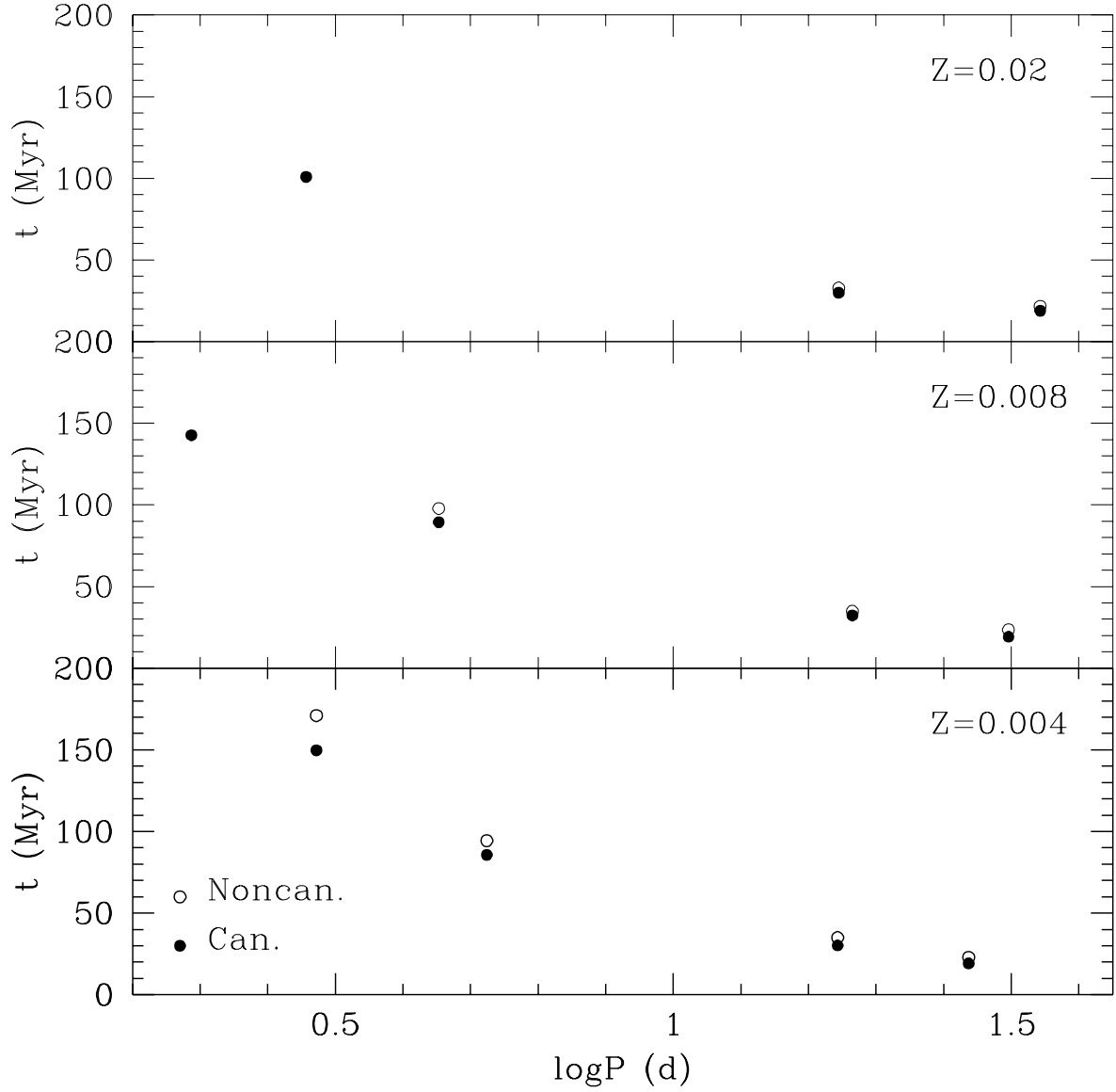}
\caption{Evolutionary ages (Myr) as a function of pulsation period (days) 
for different chemical compositions (see labeled values). Circles refer to 
fundamental periods estimated according to evolutionary predictions that 
either neglect (Canonical, filled) or account (Noncanonical, open) for 
convective core overshooting.}
\end{figure}

% Fig. 5 
\clearpage 
\begin{figure}
%\epsscale{0.80}
\plotone{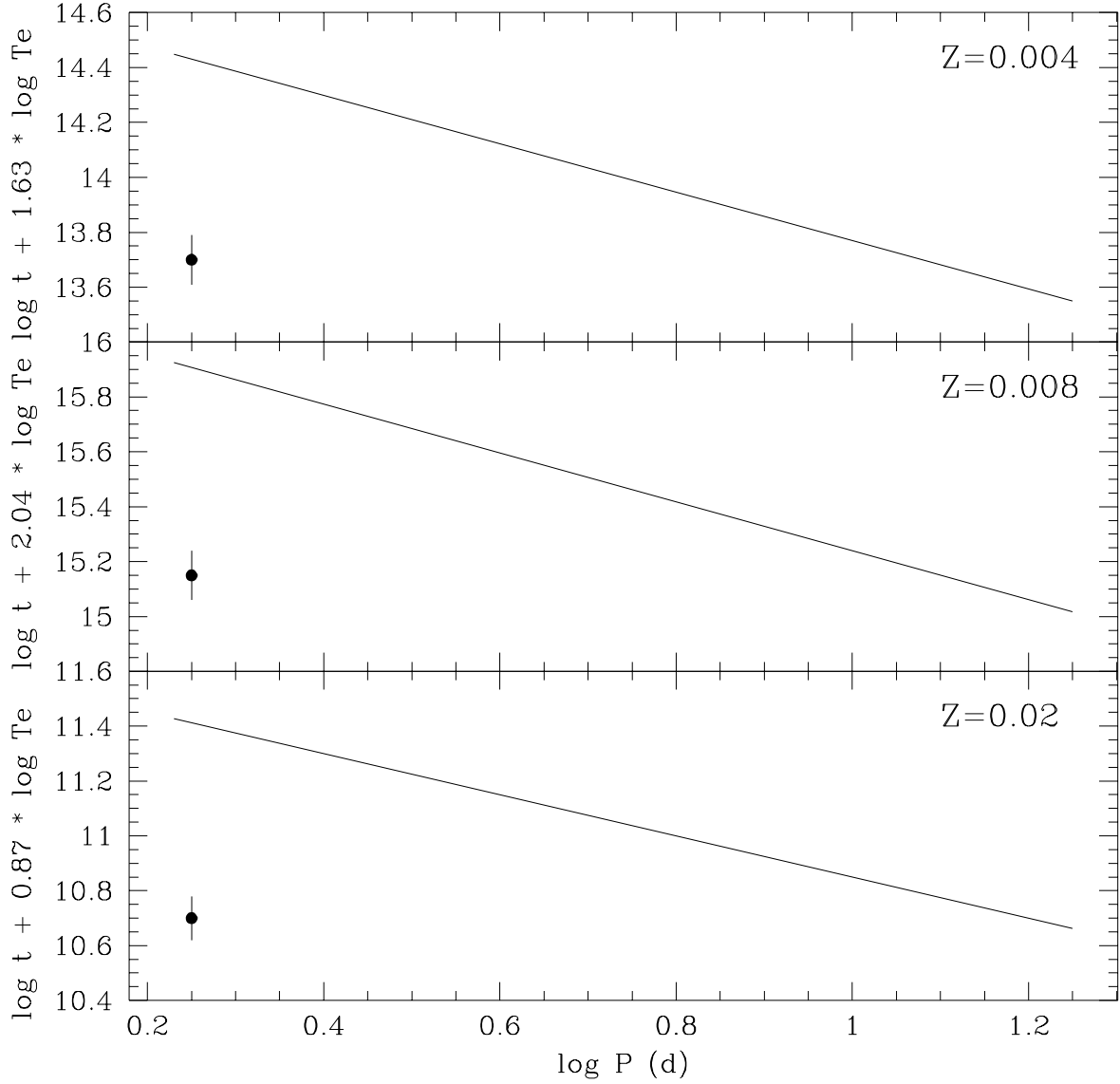}
\caption{Projection onto a plane of Period-Age-Color relations 
for fundamental Cepheids at different chemical compositions (see 
labeled values).\label{fnolin}}
\end{figure}

% Fig. 6a  
\clearpage 
\begin{figure}
%\epsscale{0.80}
\figurenum{6a}
\plotone{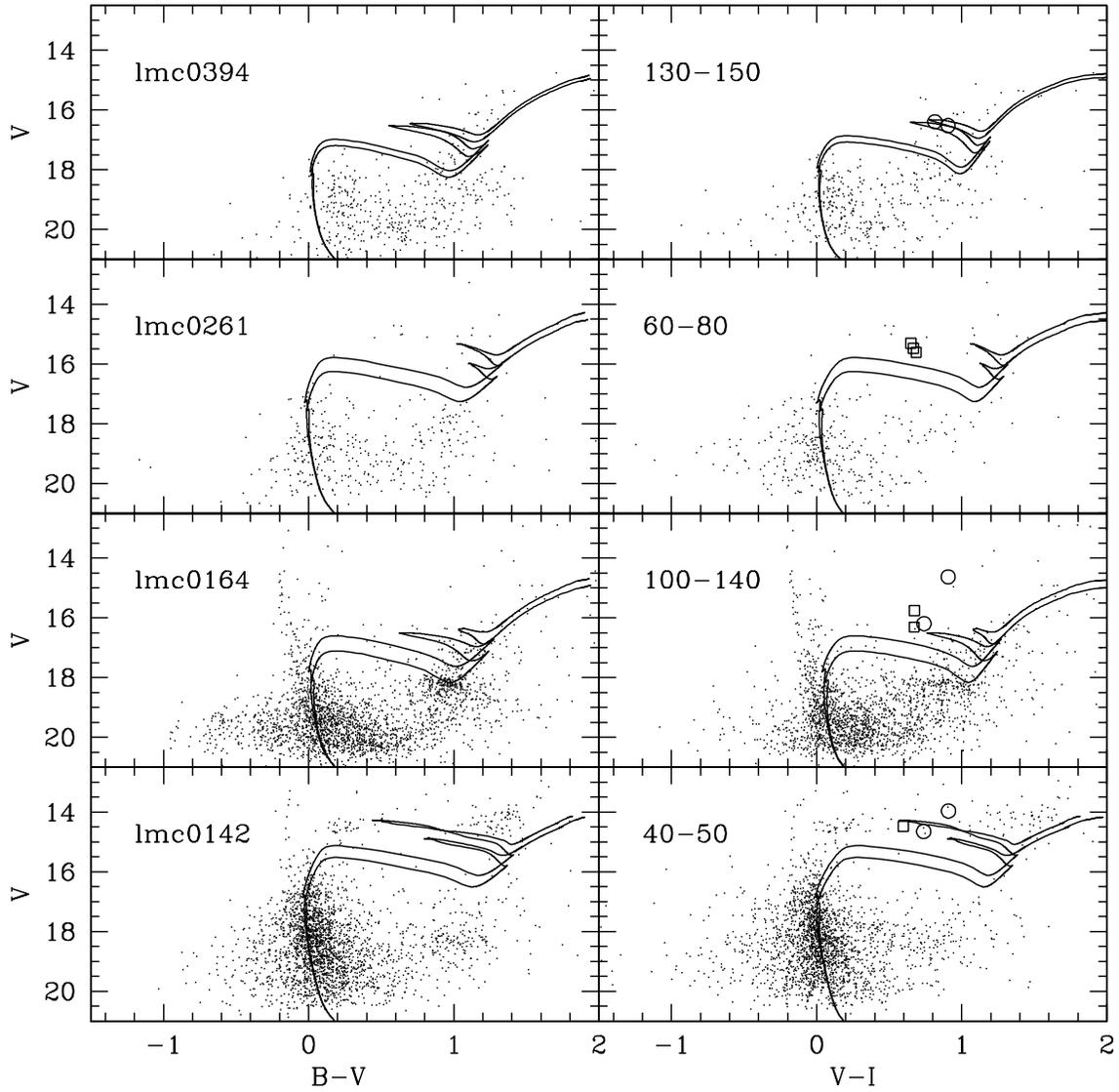}
\caption{Fit of empirical Color-Magnitude Diagrams in V,B-V 
(left)  and in V,V-I (right) of LMC clusters with stellar isochrones 
at fixed metal content (Z=0.01). Open circles and squares mark fundamental 
and first overtone cluster Cepheids respectively. Individual 
cluster ages, and reddening values are listed in Table 6.\label{fnolin}}
\end{figure}

% Fig. 6b  
\clearpage 
\begin{figure}
%\epsscale{0.80}
\figurenum{6b}
\plotone{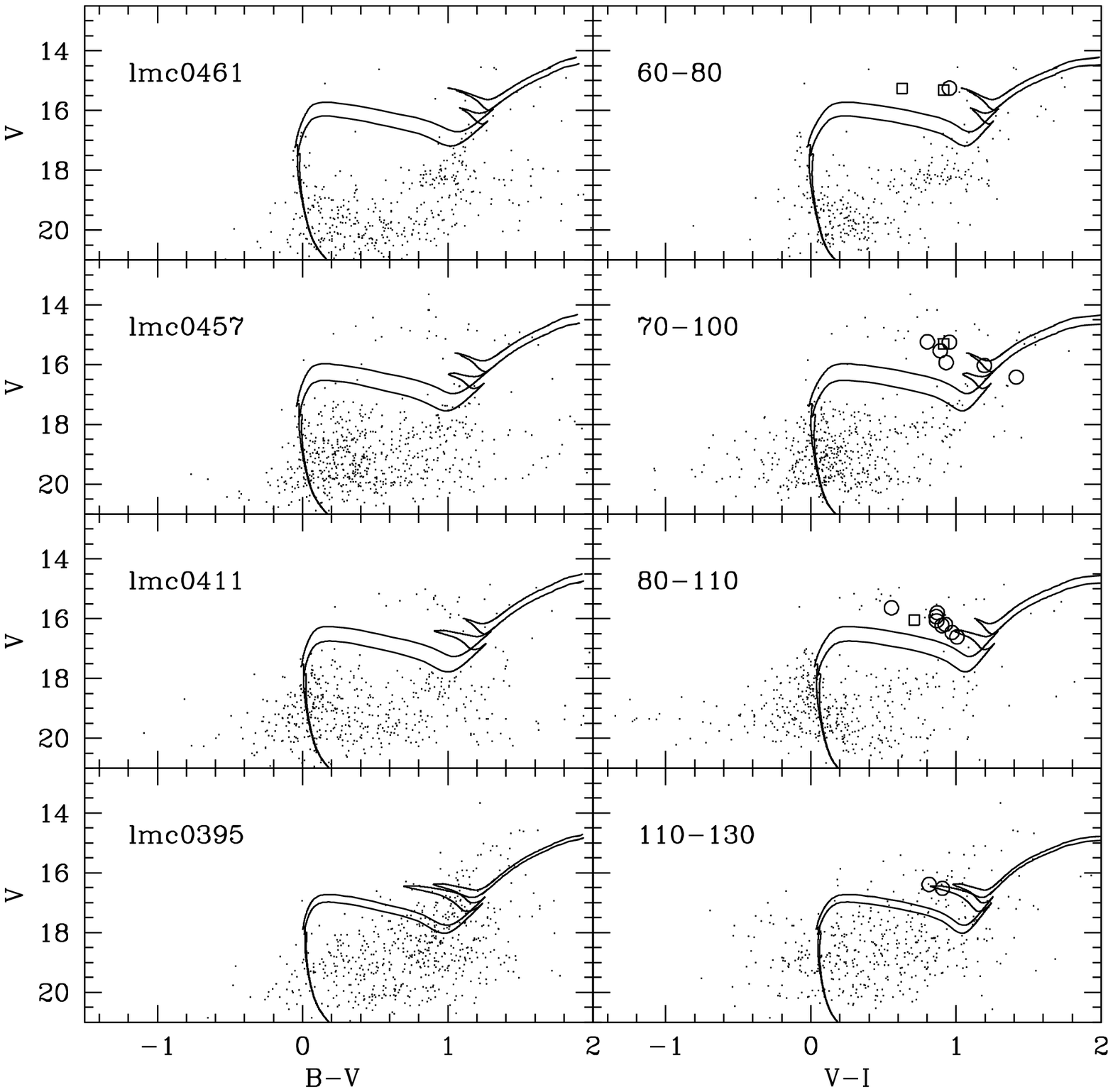}
\caption{Same as Fig. 6a but for a different sample of LMC clusters.\label{fnolin}}
\end{figure}

% Fig. 6c  
\clearpage 
\begin{figure}
%\epsscale{0.80}
\figurenum{6c}
\plotone{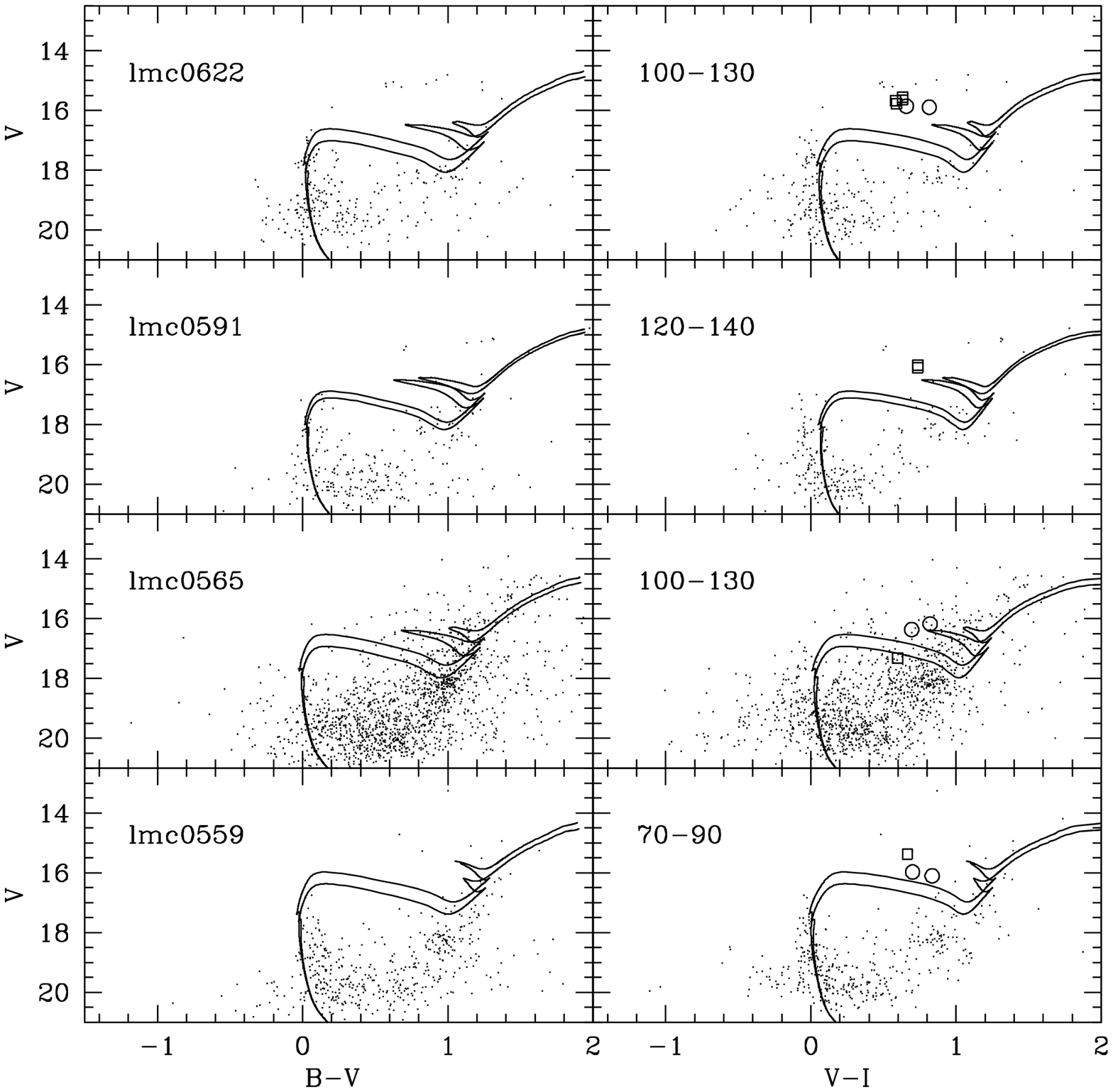}
\caption{Same as Fig. 6a but for a different sample of LMC clusters.\label{fnolin}}
\end{figure}

% Fig. 6d  
\clearpage 
\begin{figure}
%\epsscale{0.80}
\figurenum{6d}
\plotone{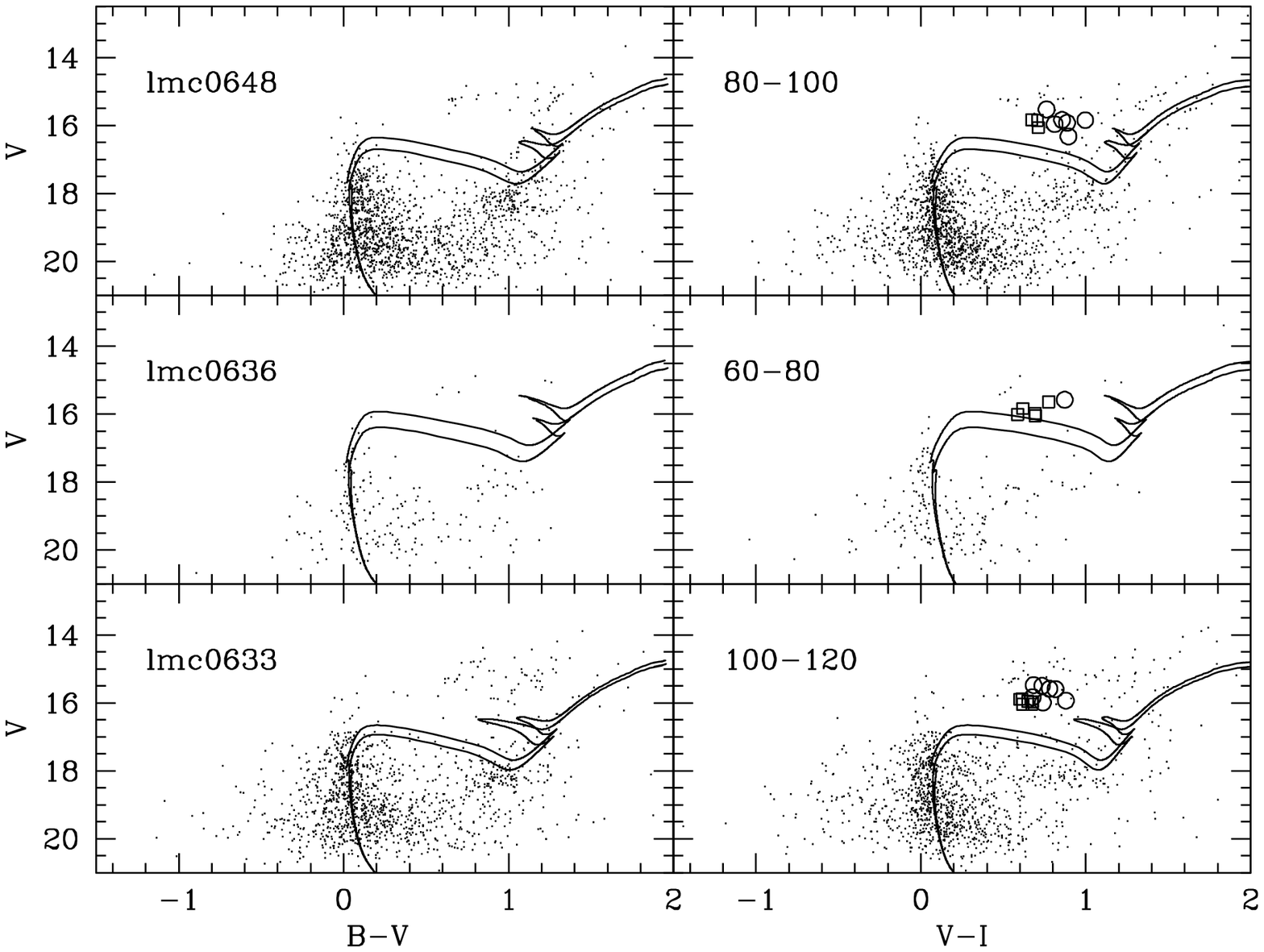}
\caption{Same as Fig. 6a but for a different sample of LMC clusters.\label{fnolin}}
\end{figure}

% Fig. 7a  
\clearpage 
\begin{figure}
%\epsscale{0.80}
\figurenum{7a}
\plotone{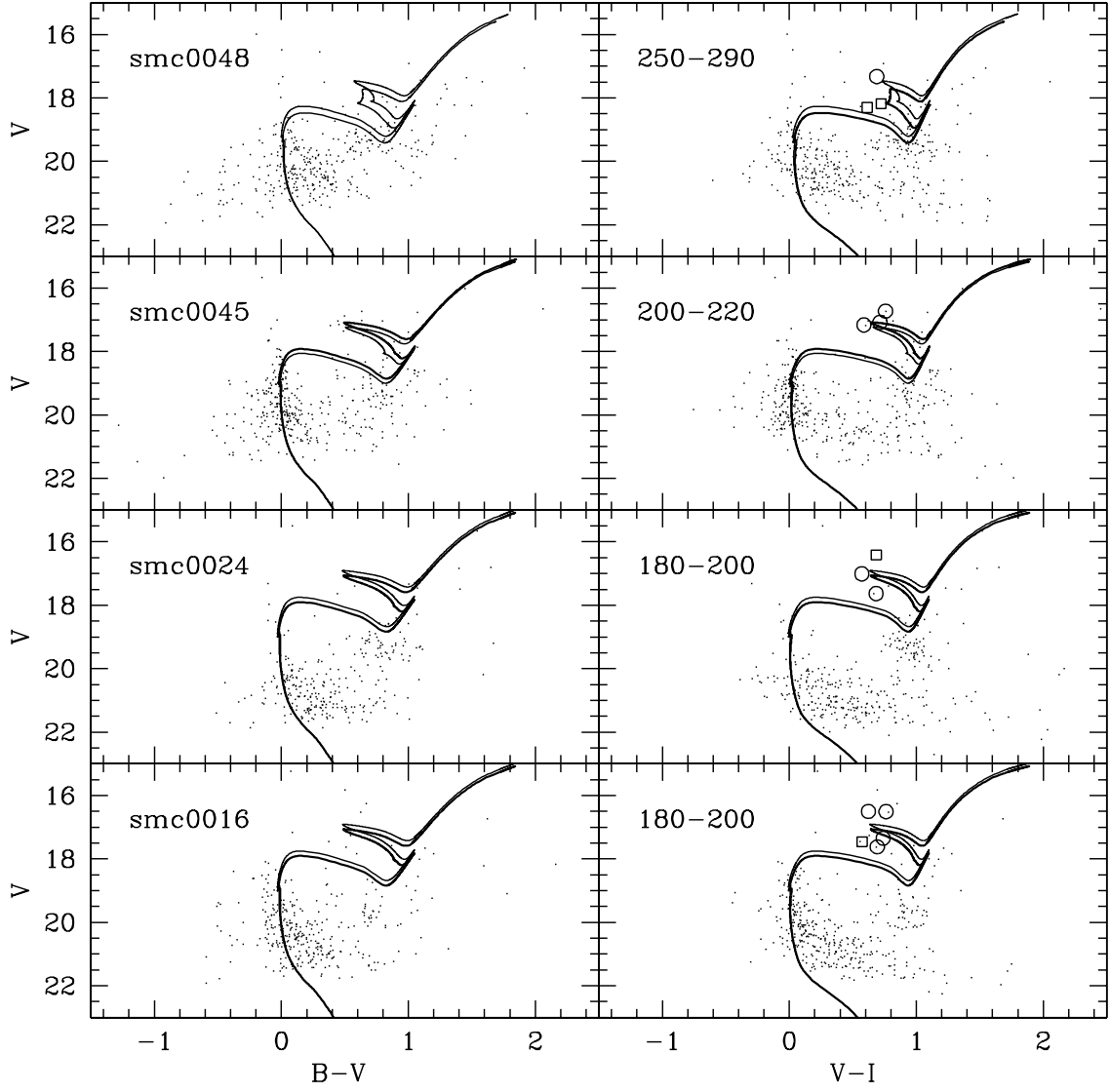}
\caption{Same as Fig. 6a but for a sample of SMC clusters and 
stellar isochrones with a metal content of Z=0.004.\label{fnolin}}
\end{figure}

% Fig. 7b  
\clearpage 
\begin{figure}
%\epsscale{0.80}
\figurenum{7b}
\plotone{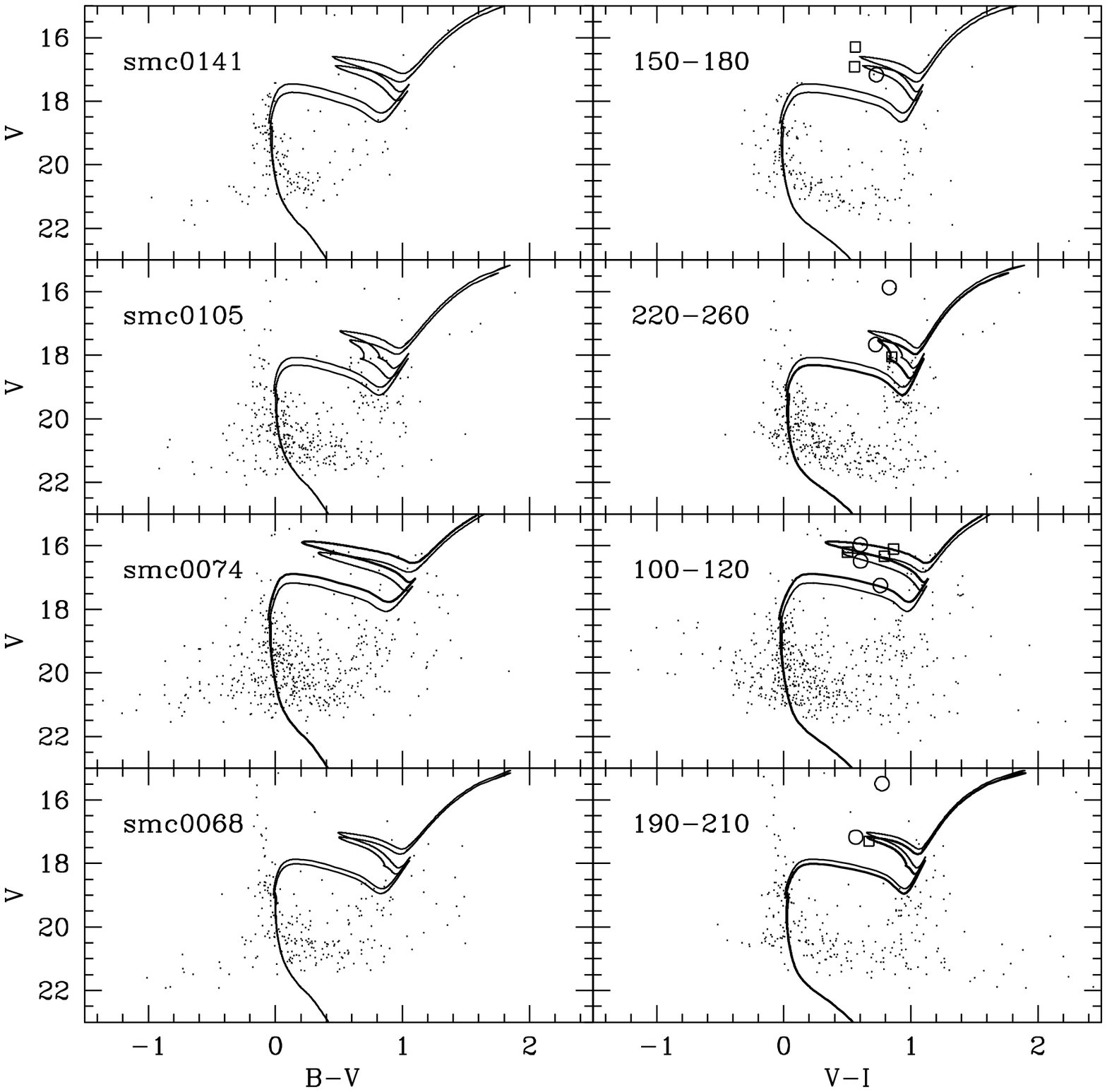}
\caption{Same as Fig. 7a but for a different sample of SMC clusters.\label{fnolin}}
\end{figure}

% Fig. 7c  
\clearpage 
\begin{figure}
%\epsscale{0.80}
\figurenum{7c}
\plotone{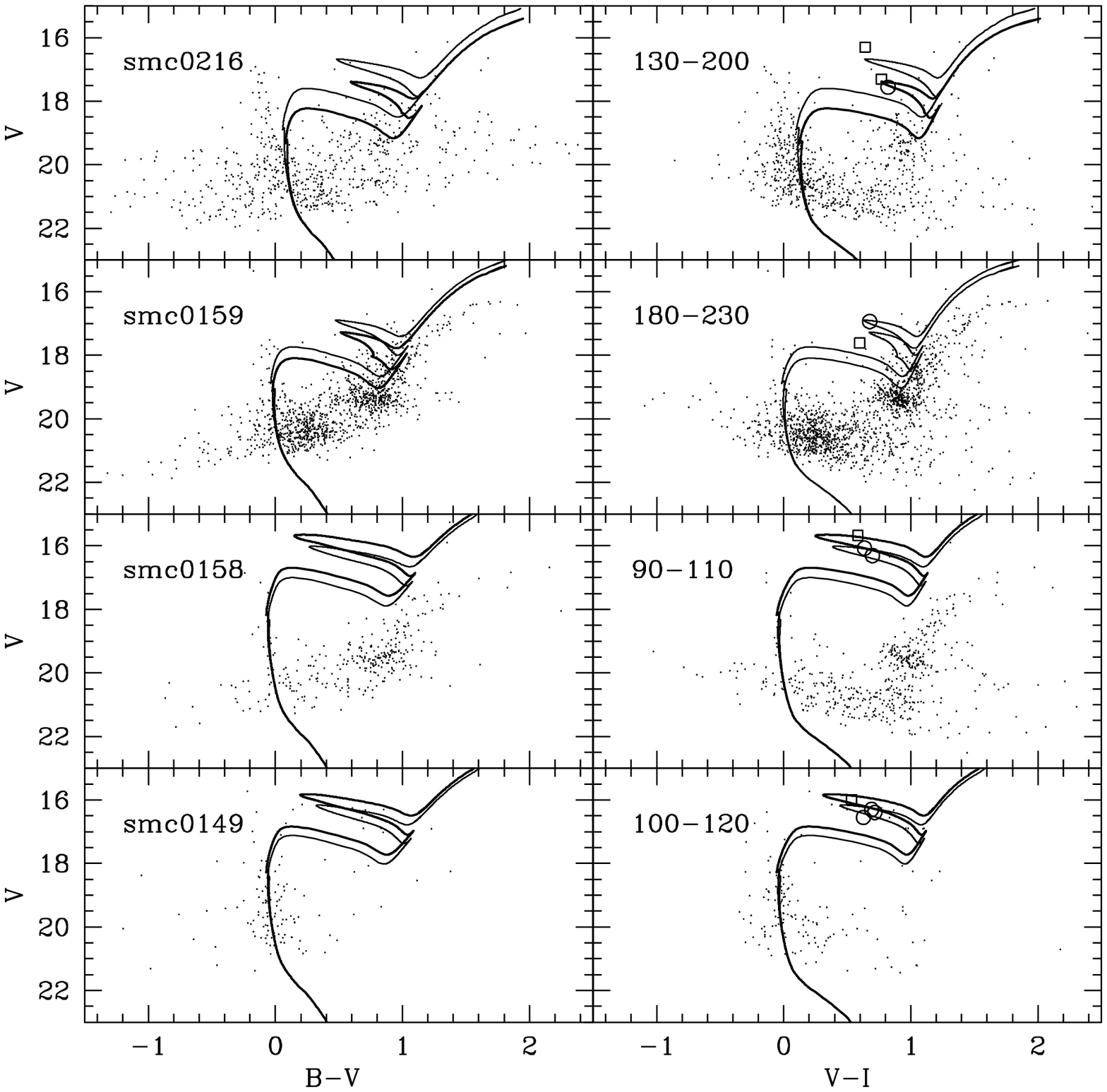}
\caption{Same as Fig. 7a but for a different sample of SMC clusters.\label{fnolin}}
\end{figure}

% Fig. 8 
\clearpage 
\begin{figure}
%\epsscale{0.80}
\figurenum{8}
\plotone{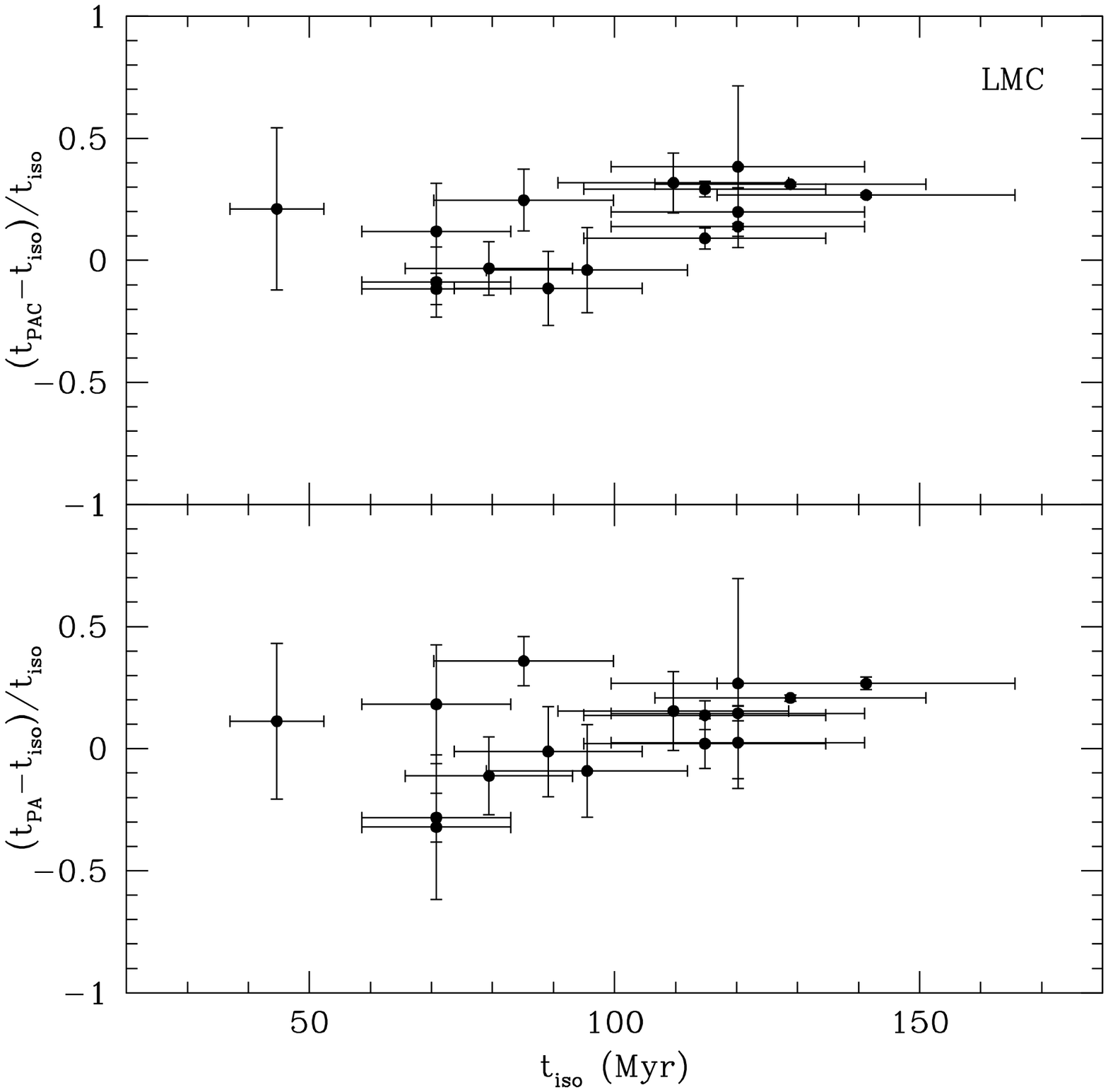}
\caption{Relative difference between evolutionary and pulsation cluster 
ages as a function of isochrone ages for LMC clusters that host at least 
two Cepheids. The top panel shows the difference with the PAC relation, 
while the bottom one with the PA relation.\label{fnolin}}
\end{figure}

% Fig. 9 
\clearpage 
\begin{figure}
%\epsscale{0.80}
\figurenum{9}
\plotone{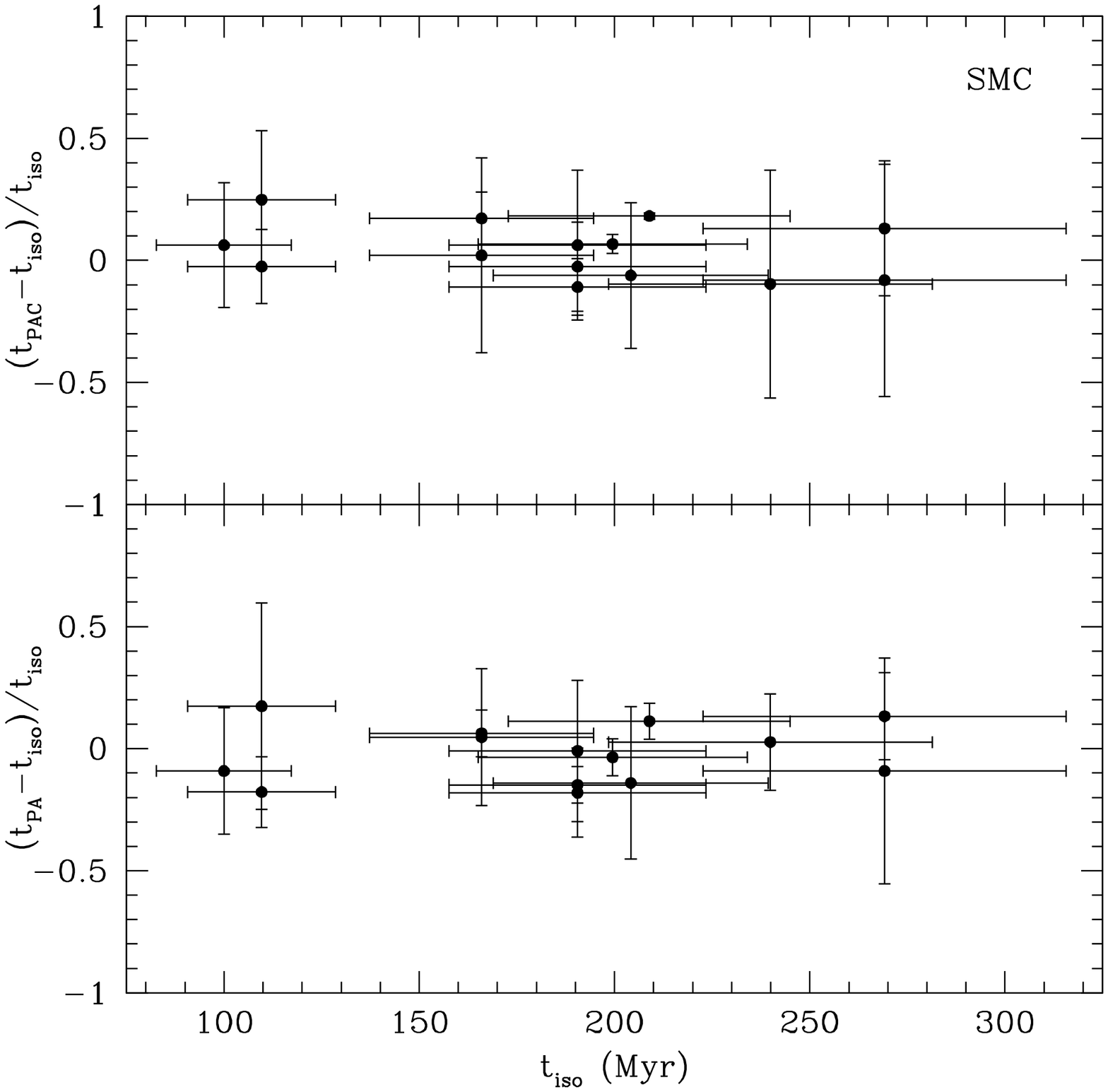}
\caption{Same as Fig. 8, but for SMC clusters. See text for more details.\label{fnolin}}
\end{figure}

% Fig. 8   
\clearpage 
\begin{figure}
\epsscale{0.90}
\figurenum{10}
\plotone{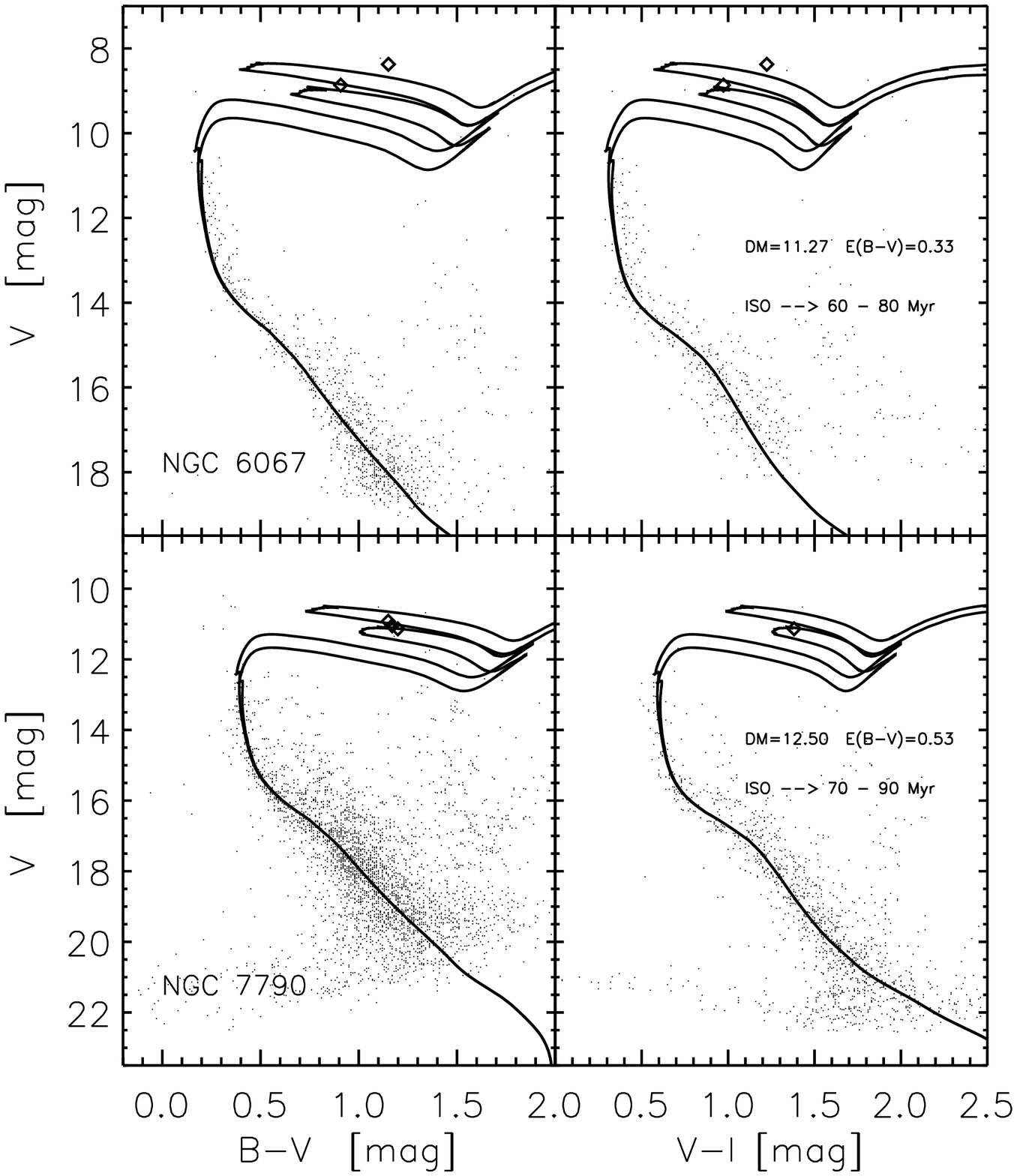}
\vspace*{1.5truecm} 
\caption{Same as Fig. 6a, but for two Galactic clusters and stellar 
isochrones with a metal content of Z=0.02.\label{fnolin}}
\end{figure}

%1========================== Table X ================================
\clearpage 
\begin{deluxetable}{cccccccccccc}
\scriptsize 
\tablecolumns{12} 
\tablewidth{0pt} 
\tablecaption{Input parameters of the pulsation models adopted to 
compute the PA and the PAC relations.\label{t2}}
\tablehead{
\colhead{$M$\tablenotemark{a}}&
\colhead{$\log L$\tablenotemark{b}} &  
\colhead{$M$\tablenotemark{a}}&
\colhead{$\log L$\tablenotemark{b}} & 
\colhead{$M$\tablenotemark{a}}&
\colhead{$\log L$\tablenotemark{b}} &  
\colhead{$M$\tablenotemark{a}}&
\colhead{$\log L$\tablenotemark{b}} &  
\colhead{$M$\tablenotemark{a}}&
\colhead{$\log L$\tablenotemark{b}} & 
\colhead{$M$\tablenotemark{a}}&
\colhead{$\log L$\tablenotemark{b}} \\ 
\colhead{(1)}&
\colhead{(2)}&
\colhead{(3)}&
\colhead{(4)}&
\colhead{(5)}&
\colhead{(6)}& 
\colhead{(7)}&
\colhead{(8)}&
\colhead{(9)}&
\colhead{(10)}&
\colhead{(11)}&
\colhead{(12)}}
\startdata
\multicolumn{2}{c}{Z=0.004\tablenotemark{c}} & \multicolumn{2}{c}{Z=0.008\tablenotemark{c}} & \multicolumn{2}{c}{Z=0.02\tablenotemark{c}} & 
\multicolumn{2}{c}{Z=0.004\tablenotemark{c}} & \multicolumn{2}{c}{Z=0.008\tablenotemark{c}} & \multicolumn{2}{c}{Z=0.02\tablenotemark{c}}\\
\multicolumn{6}{c}{First-overtone Canonical\tablenotemark{d}} & \multicolumn{6}{c}{First-overtone Noncanonical\tablenotemark{e}}\\
 3.25  & 2.49 & 3.25  &  2.45 &  3.50 & 2.51  &  3.00 & 2.62 & 3.00  & 2.58   &  3.00 & 2.52 \\ 

 3.50  & 2.61 & 3.50  &  2.47 &  4.00 & 2.72  &  3.50 & 2.86 & 3.50  &  2.82  &  4.00 & 2.97 \\

 3.80  & 2.74 & 3.80  &  2.70 &  4.50 & 2.90  &  4.00 & 3.07 & 4.00  &  3.03  &  4.60 & 3.19 \\

  4.00 & 2.82 & 4.00  &  2.78 &  5.00 & 3.07  &  5.00 & 3.30 & 5.00  &  3.30  &  4.75 & 3.24 \\

  5.00 & 3.07 & 5.00  &  3.07 &  5.50 & 3.22  &  5.25 & 3.50 & 5.25  &  3.46  & \ldots& \ldots\\

  5.50 & 3.32 & 7.00  &  3.65 &  5.60 & 3.25  &  5.50 & 3.57 & 5.50  &  3.53  & \ldots& \ldots\\

  5.80 & 3.40 & 7.30  &  3.72 &\ldots &\ldots &  5.60 & 3.60 & 5.60  &  3.55  & \ldots& \ldots\\

  7.00 & 3.65 & \ldots & \ldots&\ldots &\ldots& \ldots & \ldots & \ldots&\ldots &\ldots& \ldots \\

  7.15 & 3.73 & \ldots & \ldots&\ldots &\ldots& \ldots & \ldots & \ldots&\ldots &\ldots& \ldots \\

  7.30 & 3.76 & \ldots & \ldots&\ldots &\ldots& \ldots & \ldots & \ldots&\ldots &\ldots& \ldots\\
\multicolumn{6}{c}{Fundamental Canonical\tablenotemark{d}} & \multicolumn{6}{c}{Fundamental Noncanonical\tablenotemark{e}}\\			
 3.25 &  2.49 & 3.25 & 2.45  & 4.50  &  2.90  &  3.50  &  2.86  & 4.0  & 3.07 & 4.00  & 3.22 \\

 3.50 &  2.61 & 3.50 & 2.57  & 5.00  &  3.07  &  5.00 &  3.30  & 5.00 & 3.30 & 5.00  & 3.30\\

 3.80 &  2.74 & 3.80 & 2.70  & 6.25  &  3.42  &  7.00 &  3.85  & 7.00 & 3.85 & 7.00  & 3.85 \\

 4.00 &  2.82 & 4.00 & 2.78  & 6.50  &  3.48  &  9.00 &  4.25  & 9.00 & 4.25 & 9.00  & 4.25 \\

 5.00 &  3.07 & 5.00 & 3.07  & 6.75  &  3.54  &  11.00&  4.65  & 11.00& 4.65 & 11.00 & 4.65\\

 7.00 &  3.65 & 6.55 & 3.55  & 7.00  &  3.65  & \ldots& \ldots & \ldots& \ldots & \ldots& \ldots \\

 7.15 &  3.73 & 6.70 & 3.59  & 9.00  &  4.00  & \ldots& \ldots & \ldots& \ldots & \ldots& \ldots \\

 7.30 &  3.76 & 6.85 & 3.62  & 11.00 &  4.40 & \ldots& \ldots & \ldots& \ldots & \ldots& \ldots \\

 7.45 &  3.79 & 7.00 & 3.65  & \ldots& \ldots & \ldots& \ldots & \ldots& \ldots & \ldots& \ldots \\

 9.00 &  4.00 & 7.15 & 3.69  & \ldots& \ldots & \ldots& \ldots & \ldots& \ldots & \ldots& \ldots \\

11.00 &  4.40 & 7.30 & 3.72  & \ldots& \ldots & \ldots& \ldots & \ldots& \ldots & \ldots& \ldots \\
\ldots& \ldots& 7.45 & 3.75  & \ldots& \ldots & \ldots& \ldots & \ldots& \ldots & \ldots& \ldots \\  
\ldots& \ldots& 9.00 & 4.00  & \ldots& \ldots & \ldots& \ldots & \ldots& \ldots & \ldots& \ldots \\  
\ldots& \ldots& 11.00& 4.40  & \ldots& \ldots & \ldots& \ldots & \ldots& \ldots & \ldots& \ldots \\  
\enddata
\tablenotetext{a}{Stellar mass (solar units).
\hspace*{2.5mm} $^b$Logarithmic luminosity (solar units).
\hspace*{2.5mm} $^c$Metal (Z) abundances by mass; the helium abundances 
are Y=0.25 (Z=0.004, Z=0.008) and Y=0.28 (Z=0.02).
\hspace*{2.5mm} $^d$Pulsation models constructed by adopting a M/L 
relation based on evolutionary models that neglect the convective 
core overshooting (Bono et al. 1999).  
\hspace*{2.5mm} $^e$Pulsation models constructed by adopting a M/L 
relation that mimics the behavior of evolutionary models that account 
for the convective core overshooting during H-burning 
(Chiosi et al. 2003; Bono et al. 1999).}
\end{deluxetable}

%2========================== Table X ================================
\clearpage 
\begin{deluxetable}{ccccccc}
\normalsize
%\scriptsize 
\tablecolumns{7} 
\tablewidth{0pt} 
\tablecaption{Analytical relations 
($\log T_e=\alpha + \beta \log L/L_{\odot}$) for the boundaries of the 
instability strip at different chemical compositions. \label{t2}}
\tablehead{
\colhead{$Z$\tablenotemark{a}}&
\colhead{$\alpha$\tablenotemark{b}}&
\colhead{$\beta$\tablenotemark{c}}&
\colhead{$\sigma$\tablenotemark{d}}& 
\colhead{$\alpha$\tablenotemark{b}}&
\colhead{$\beta$\tablenotemark{c}}&
\colhead{$\sigma$\tablenotemark{d}} \\ 
\colhead{(1)}&
\colhead{(2)}&
\colhead{(3)}&
\colhead{(4)}&
\colhead{(5)}&
\colhead{(6)}&
\colhead{(7)}}
\startdata
      &\multicolumn{3}{c}{Canonical\tablenotemark{e}} & \multicolumn{3}{c}{Noncanonical\tablenotemark{f}} \\
      &\multicolumn{6}{c}{First-overtone blue edge}\\
0.004 & 3.933$\pm$0.008 & -0.041$\pm$0.005  & 0.008 & 3.967$\pm$0.002 & -0.053$\pm$0.003  & 0.002 \\
0.008 & 3.961$\pm$0.005 & -0.052$\pm$0.004  & 0.005 & 3.959$\pm$0.002 & -0.052$\pm$0.002  & 0.002 \\
0.02  & 3.974$\pm$0.009 & -0.061$\pm$0.014  & 0.008 &\ldots &\ldots &\ldots  \\
      &\multicolumn{6}{c}{First-overtone red edge}\\
0.004 & 3.820$\pm$0.008 & -0.017$\pm$0.005  & 0.007 &  3.869$\pm$0.011  & -0.031$\pm$0.012 & 0.010 \\
0.008 & 3.831$\pm$0.006 & -0.021$\pm$0.005  & 0.005 & 3.839$\pm$0.010   & -0.023$\pm$0.010& 0.009 \\
0.02  & 3.803$\pm$0.003 & -0.012$\pm$0.005  & 0.003 &\ldots &\ldots &\ldots \\
      &\multicolumn{6}{c}{Fundamental blue edge}\\			
0.004 & 3.812$\pm$0.007 & -0.012$\pm$0.003  & 0.07 & 3.856$\pm$0.008   & -0.023$\pm$0.006 & 0.007 \\
0.008 & 3.838$\pm$0.010 & -0.021$\pm$0.005   &  0.010  & 3.907$\pm$0.009   & -0.040$\pm$0.007& 0.008\\
0.02  & 3.949$\pm$0.007 & -0.059$\pm$0.006  &   0.007   &  3.955$\pm$0.007 & -0.059$\pm$0.006& 0.007\\
0.02\tablenotemark{g}& 3.940$\pm$0.005   & -0.056$\pm$0.005  & 0.005  &\ldots &\ldots &\ldots \\
      &\multicolumn{6}{c}{Fundamental red edge}\\			
0.004 & 3.923$\pm$0.008 & -0.061$\pm$0.004   & 0.008  & 3.957$\pm$0.007   & -0.072$\pm$0.006& 0.006 \\
0.008 & 3.947$\pm$0.008 & -0.070$\pm$0.004   &  0.008 & 3.980$\pm$0.012   & -0.082$\pm$0.011&  0.010\\
0.02  & 4.039$\pm$0.007 & -0.101$\pm$0.005  &   0.007 &  4.118$\pm$0.014  &  -0.122$\pm$0.012& 0.012\\
0.02\tablenotemark{g} & 4.061$\pm$0.009   & -0.110$\pm$0.008  &   0.008  &\ldots &\ldots &\ldots  \\
\enddata
\tablenotetext{a}{Metal abundance. 
\hspace*{2.5mm} $^b$ Zero Point.
\hspace*{2.5mm} $^c$ Slope.
\hspace*{2.5mm} $^d$ Standard deviations.
\hspace*{2.5mm} $^e$Pulsation models constructed by adopting a M/L 
relation based on evolutionary models that neglect the convective 
core overshooting (Bono et al. 1999).  
\hspace*{2.5mm} $^d$Pulsation models constructed by adopting a M/L 
relation that mimics the behavior of behavior of models that account 
for the convective core overshooting during H-burning 
(Chiosi et al. 2003; Bono et al. 1999). 
\hspace*{2.5mm} $^g$ Pulsation models from Petroni et al. (2003).}  
\end{deluxetable}

%2========================== Table X ================================
\clearpage 
\begin{deluxetable}{ccccc}
\normalsize
%\scriptsize 
\tablecolumns{7} 
\tablewidth{0pt} 
\tablecaption{Analytical relations 
($\log T_e=\alpha + \beta \log L/L_{\odot} + \gamma (\log L/L_{\odot})^2 $) for the 
boundaries of the canonical fundamental instability strip at different chemical 
compositions.\label{t2}}
\tablehead{
\colhead{$Z$\tablenotemark{a}}&
\colhead{$\alpha$\tablenotemark{b}}&
\colhead{$\beta$\tablenotemark{c}}&
\colhead{$\gamma$\tablenotemark{d}}&
\colhead{$\sigma$\tablenotemark{e}} \\ 
\colhead{(1)}&
\colhead{(2)}&
\colhead{(3)}&
\colhead{(4)}&
\colhead{(5)}}
\startdata
      \multicolumn{5}{c}{Blue edge}\\			
0.004 & 3.617$\pm$0.047 & 0.108$\pm$0.029 & -0.018$\pm$0.004 & 0.001 \\
0.008 & 3.557$\pm$0.050 & 0.153$\pm$0.031 & -0.026$\pm$0.005 & 0.001 \\
0.02  & 3.729$\pm$0.122 & 0.064$\pm$0.068 & -0.017$\pm$0.009 & 0.001 \\
0.02\tablenotemark{f}& 3.646$\pm$0.105 & 0.102$\pm$0.056 & -0.021$\pm$0.007 & 0.001\\
      \multicolumn{5}{c}{Red edge}\\			
0.004 & 3.678$\pm$0.050 & 0.090$\pm$0.026 & -0.026$\pm$0.005 & 0.001 \\
0.008 & 3.709$\pm$0.028 & 0.077$\pm$0.017 & -0.022$\pm$0.003 & 0.001 \\
0.02  & 3.763$\pm$0.082 & 0.053$\pm$0.045 & -0.021$\pm$0.006 & 0.001 \\
0.02\tablenotemark{f}& 3.509$\pm$0.056 & 0.187$\pm$0.030 & -0.039$\pm$0.004 & 0.001\\
\enddata
\tablenotetext{a}{Metal abundance. 
\hspace*{2.5mm} $^b$ Zero Point.
\hspace*{2.5mm} $^c$ Coefficient of the linear luminosity term.
\hspace*{2.5mm} $^d$ Coefficient of the quadratic luminosity term.
\hspace*{2.5mm} $^e$ Standard deviation.
\hspace*{2.5mm} $^f$ Pulsation models from Petroni et al. (2003).}  
\end{deluxetable}

%3========================== Table X ================================
\clearpage 
\begin{deluxetable}{cccc}
%\scriptsize 
\tablecolumns{4} 
\tablewidth{0pt} 
\tablecaption{Period-Age relations 
($\log t\tablenotemark{a}=\alpha + \beta \log P$) at different chemical 
compositions. \label{t2}}
\tablehead{
\colhead{$Z$\tablenotemark{b}}&
\colhead{$\alpha$\tablenotemark{c}}&
\colhead{$\beta$\tablenotemark{d}}&
\colhead{$\sigma$\tablenotemark{e}} \\ 
\colhead{(1)}&
\colhead{(2)}&
\colhead{(3)}&
\colhead{(4)}}
\startdata
\multicolumn{4}{c}{First-overtone}\\
0.004 &  8.41$\pm$0.07  & -1.07$\pm$0.02  & 0.07   \\
0.01  & 8.29$\pm$0.08   & -0.80$\pm$0.03  & 0.08   \\
0.02  & 8.08$\pm$0.04   & -0.39$\pm$0.04  & 0.04   \\
\multicolumn{4}{c}{Fundamental}\\			
0.004 & 8.49$\pm$0.09   & -0.79$\pm$0.01  & 0.09   \\
0.01  & 8.41$\pm$0.10   & -0.78$\pm$0.01  & 0.10   \\
0.02  & 8.31$\pm$0.08   & -0.67$\pm$0.01  & 0.08   \\
\enddata
\tablenotetext{a}{Cepheid age in years.   
\hspace*{2.5mm} $^b$Metal abundance.   
\hspace*{2.5mm} $^c$Zero Point.
\hspace*{2.5mm} $^d$Slope.
\hspace*{2.5mm} $^e$Standard deviations.}  
\end{deluxetable}

%4========================== Table X ================================
\clearpage 
\begin{deluxetable}{ccccccccc}
%\scriptsize 
\tablecolumns{9} 
\tablewidth{0pt} 
\tablecaption{Period and ages for selected evolutionary models. 
\label{t2}}
\tablehead{
\colhead{$M/M_\odot$\tablenotemark{a}}&
\colhead{$\log L/L_\odot$\tablenotemark{b}}&
\colhead{$\log T_e$\tablenotemark{c}}&
\colhead{$\log P$\tablenotemark{d}}&
\colhead{$Age_{ISO}$\tablenotemark{e}}&
\colhead{$\log L/L_\odot$\tablenotemark{b}}&
\colhead{$\log T_e$\tablenotemark{c}}&
\colhead{$\log P$\tablenotemark{d}}&
\colhead{$Age_{ISO}$\tablenotemark{e}} \\ 
\colhead{(1)}&
\colhead{(2)}&
\colhead{(3)}&
\colhead{(4)}&
\colhead{(5)}&
\colhead{(6)}&
\colhead{(7)}&
\colhead{(8)}&
\colhead{(9)}}
\startdata
 & \multicolumn{4}{c}{Canonical}&\multicolumn{4}{c}{Noncanonical}\\
\multicolumn{9}{c}{Z=0.004, Y=0.251\tablenotemark{f}}\\			
 4& 2.935 & 3.760& 0.472&  149.69& 3.149& 3.757& 0.682&  171.04\\
 5& 3.249 & 3.749& 0.724&   85.65& 3.419& 3.744& 0.897&   94.35\\
 8& 3.900 & 3.725& 1.243&   30.16& 4.089& 3.712& 1.462&   34.98\\
10& 4.159 & 3.716& 1.437&   19.16& 4.382& 3.698& 1.702&   22.96\\
\multicolumn{9}{c}{Z=0.008, Y=0.256}\\			
 4& 2.751 & 3.767& 0.287&  142.73&\ldots&\ldots&\ldots& \ldots\\
 5& 3.162 & 3.749& 0.653&   89.39& 3.360& 3.739& 0.870&   97.85\\
 8& 3.879 & 3.716& 1.265&   32.36& 4.049& 3.697& 1.488&   34.86\\
10& 4.163 & 3.703& 1.496&   19.24& 4.354& 3.678& 1.756&   23.56\\
\multicolumn{9}{c}{Z=0.0198, Y=0.273}\\			
 5& 2.967 & 3.757& 0.457& 100.92  &\ldots &\ldots &\ldots& \ldots \\
 8& 3.760 & 3.693& 1.245& 29.93 & 3.951 & 3.679& 1.469& 32.79 \\
10& 4.073 & 3.668& 1.543& 18.89 & 4.296& 3.648& 1.817& 21.63 \\
\enddata
\tablenotetext{a}{Stellar mass (solar units).   
\hspace*{2.5mm} $^b$Luminosity (solar units). 
\hspace*{2.5mm} $^c$Effective temperature (K).
\hspace*{2.5mm} $^d$Period (days).
\hspace*{2.5mm} $^e$Stellar age (My).   
\hspace*{2.5mm} $^f$Metal (Z) and helium (Y) abundances by mass adopted 
to construct evolutionary models.}  
\end{deluxetable}

%5========================== Table Y ================================
\clearpage 
\begin{deluxetable}{ccccc}
%\scriptsize 
\tablecolumns{5} 
\tablewidth{0pt} 
\tablecaption{Period-Age-Color relations 
($\log t\tablenotemark{a}=\alpha + \beta \log P + \gamma CI$) at different 
chemical compositions. \label{t2}}
\tablehead{
\colhead{$Z$\tablenotemark{b}}&
\colhead{$\alpha$\tablenotemark{c}}&
\colhead{$\beta$\tablenotemark{d}}&
\colhead{$\gamma$\tablenotemark{e}}&
\colhead{$\sigma$\tablenotemark{f}} \\ 
\colhead{(1)}&
\colhead{(2)}&
\colhead{(3)}&
\colhead{(4)}&
\colhead{(5)}}
\startdata
\multicolumn{5}{c}{First-overtone}\\
0.004 & 18.68$\pm$0.06   & -1.184$\pm$0.02  & -2.70$\pm$0.33\tablenotemark{g}    &  0.06    \\
0.004 &  8.14$\pm$0.06  & -1.17$\pm$0.02  & 0.65$\pm$0.08\tablenotemark{h}    & 0.06     \\
0.004 &  8.06$\pm$0.06  & -1.16$\pm$0.02  & 0.64$\pm$0.08\tablenotemark{i}    & 0.06     \\
0.01  &  18.27$\pm$0.08  & -0.89$\pm$0.04  & -2.63$\pm$0.61\tablenotemark{g}    &    0.08   \\
0.01  & 8.04$\pm$0.08   & -0.89$\pm$0.04  & 0.53$\pm$0.14\tablenotemark{h}    & 0.08      \\
0.01  & 7.94$\pm$0.08   & -0.88$\pm$0.03  & 0.61$\pm$0.15\tablenotemark{i}    &  0.08     \\
0.02  & 14.15$\pm$0.04   & -0.48$\pm$0.05  & -1.60$\pm$0.60\tablenotemark{g}    &  0.04      \\
0.02  & 7.94$\pm$0.04   & -0.47$\pm$0.05  & 0.31$\pm$0.12\tablenotemark{h}    &    0.04   \\
0.02  & 7.89$\pm$0.04   & -0.46$\pm$0.05  & 0.36$\pm$0.14\tablenotemark{i}    &    0.04   \\
\multicolumn{5}{c}{Fundamental}\\			
0.004 & 14.65$\pm$0.09   & -0.88$\pm$0.02  & -1.63$\pm$0.36\tablenotemark{g}    &  0.09     \\
0.004 &  8.34$\pm$0.09  & -0.90$\pm$0.02  & 0.35$\pm$0.06\tablenotemark{h}    & 0.09     \\
0.004 & 8.24$\pm$0.09   & -0.88$\pm$0.02  & 0.42$\pm$0.08\tablenotemark{i}    & 0.09     \\
0.01  &  16.13$\pm$0.09  & -0.89$\pm$0.03  & -2.04$\pm$0.42\tablenotemark{g}    &  0.09     \\
0.01  & 8.24$\pm$0.09   & -0.90$\pm$0.03  & 0.36$\pm$0.07\tablenotemark{h}    &  0.09     \\
0.01  & 8.13$\pm$0.09   & -0.89$\pm$0.03  & 0.49$\pm$0.10\tablenotemark{i}    &  0.09     \\
0.02  &  11.60$\pm$0.08  & -0.75$\pm$0.03  & -0.87$\pm$0.34\tablenotemark{g}    &   0.08    \\
0.02  &  8.25$\pm$0.08  & -0.76$\pm$0.03  & 0.16$\pm$0.06\tablenotemark{h}    &  0.08     \\
0.02  & 8.20$\pm$0.08   & -0.75$\pm$0.03  & 0.21$\pm$0.08\tablenotemark{i}    &   0.08    \\
\enddata
\tablenotetext{a}{Cepheid age in years.  
\hspace*{2.5mm} $^b$Metal abundance.  
\hspace*{2.5mm} $^c$Zero Point.
\hspace*{2.5mm} $^d$Coefficient of the period term.
\hspace*{2.5mm} $^e$Coefficient of the color term.
\hspace*{2.5mm} $^f$Standard deviations. 
\hspace*{2.5mm} $^g$The color term is $\log T_e$. 
\hspace*{2.5mm} $^h$The color term is $B-V$.
\hspace*{2.5mm} $^i$The color term is $V-I$.}
\end{deluxetable}

%4========================== Table X ================================
\clearpage 
\begin{deluxetable}{cccccccccc}
%\scriptsize 
\small  
\tablecolumns{10} 
\tablewidth{0pt} 
\tablecaption{Age difference between canonical PA and PAC relations at 
selected periods. \label{t2}}
\tablehead{
\colhead{$\log P$\tablenotemark{a}}&
\colhead{PA\tablenotemark{b}}&
\colhead{PAC(BE)\tablenotemark{c}}&
\colhead{PAC(RE)\tablenotemark{d}}&
\colhead{PA\tablenotemark{b}}&
\colhead{PAC(BE)\tablenotemark{c}}&
\colhead{PAC(RE)\tablenotemark{d}}&
\colhead{PA\tablenotemark{b}}&
\colhead{PAC(BE)\tablenotemark{c}}&
\colhead{PAC(RE)\tablenotemark{d}}\\ 
\colhead{(1)}&
\colhead{(2)}&
\colhead{(3)}&
\colhead{(4)}&
\colhead{(5)}&
\colhead{(6)}&
\colhead{(7)}&
\colhead{(8)}&
\colhead{(9)}&
\colhead{(10)}}
\startdata
 & \multicolumn{3}{c}{Z=0.004}&\multicolumn{3}{c}{Z=0.008}&\multicolumn{3}{c}{Z=0.02}\\
  1   &7.70 &7.64(3.76) &7.72(3.71) &7.63 &7.57(3.76) &7.67(3.71) &7.64 & 7.60(3.73) & 7.65(3.68) \\
  1.5 &7.30 &7.23(3.74) &7.33(3.68) &7.24 &7.21(3.72) &7.30(3.67) &7.30 & 7.27(3.68) & 7.31(3.64) \\
\enddata
\tablenotetext{a}{Selected period.   
\hspace*{2.5mm} $^b$Logarithmic age based on the PA relation. 
\hspace*{2.5mm} $^c$Logarithmic age based on the PAC relation along the blue 
edge of the instability strip. The number in parentheses is the logarithmic 
temperature in which the period is equal to $log P=1$, or $1.5$. 
\hspace*{2.5mm} $^d$Logarithmic age based on the PAC relation along the red  
edge of the instability strip.The number in parentheses is the logarithmic
temperature in which the period is equal to $log P=1$, or $1.5$.}  
\end{deluxetable}

%6========================== Table X ================================
\clearpage 
\begin{deluxetable}{ccccccccc}
\tablecolumns{9} 
\tablewidth{0pt} 
\tablecaption{Age estimates for LMC and SMC open clusters that host at 
least two Cepheids.\label{t2}}
\tablehead{
\colhead{ID\tablenotemark{a}}&
\colhead{$N_c$\tablenotemark{b}}&
\colhead{$E(B-V)$\tablenotemark{c}}&
\colhead{$\log t_{ISO}$\tablenotemark{d}}&
\colhead{$\log t_{PA}$\tablenotemark{e}}&
\colhead{$\sigma_{PA}$\tablenotemark{f}}&
\colhead{$\log t_{PAC}$\tablenotemark{g}}&
\colhead{$\sigma_{PAC}$\tablenotemark{h}} \\ 
\colhead{(1)}&
\colhead{(2)}&
\colhead{(3)}&
\colhead{(4)}&
\colhead{(5)}&
\colhead{(6)}&
\colhead{(7)}&
\colhead{(8)}} 
\startdata
LMC0394  &    2 & 0.138 & 8.15 &  8.015&  0.015& 8.015& 0.005 \\
LMC0261  &    3 & 0.142 & 7.85 &  7.958&  0.033& 7.898& 0.024 \\
LMC0164  &    4 & 0.144 & 8.08 &  7.945& 0.255& 7.870& 0.233 \\
$LMC0164$&   3  & 0.141 & 8.08 &  8.069& 0.066& 7.984& 0.054 \\
LMC0142  &   3  & 0.152 & 7.65 &  7.598& 0.156& 7.547& 0.183 \\
LMC0461  &   3  & 0.120 & 7.85 &  7.763& 0.129& 7.795& 0.097  \\
LMC0457  &   7 & 0.120 & 7.93 &  7.737& 0.068& 7.807& 0.073  \\
LMC0411  &   9  & 0.145 & 7.98 &  8.018& 0.075& 7.997& 0.073  \\
LMC0395  &   2  & 0.138 & 8.08 &  8.012& 0.015& 8.015& 0.005  \\
LMC0622  &   6  & 0.148 & 8.06 &  7.996& 0.029& 7.910& 0.019  \\
LMC0591  &   2  & 0.147 & 8.11 & 8.009&  0.006& 7.948& 0.004  \\
LMC0565  &   2  & 0.121 & 8.06 & 8.051&  0.045& 8.019& 0.020  \\
LMC0559  &   3 & 0.121 & 7.90 & 7.946&  0.062& 7.914& 0.046  \\ 
LMC0648  &   9  & 0.175 & 7.95 & 7.955&  0.079& 7.997& 0.059  \\
LMC0636  &   6  & 0.185 & 7.85 & 7.971&  0.097& 7.887& 0.057  \\
LMC0633  &   13 & 0.162 & 8.04 & 7.967&  0.083& 7.874& 0.078  \\

SMC0048  &    3  & 0.101& 8.43 &  8.468& 0.184& 8.464& 0.191 \\
$SMC0048$&    2  & 0.101& 8.43 &  8.368& 0.089& 8.369& 0.138 \\
SMC0045  &    3  & 0.094& 8.32 &  8.268& 0.036& 8.233& 0.007 \\
SMC0024  &    2  & 0.089& 8.28 &  8.340& 0.028& 8.291& 0.077 \\
SMC0016  &    5  & 0.089& 8.28 &  8.284& 0.124& 8.252& 0.142 \\
$SMC0016$&    3  & 0.089& 8.28 &  8.352& 0.067& 8.325& 0.045 \\
SMC0141  &    3  & 0.079& 8.22 &  8.192& 0.044& 8.138& 0.057 \\
SMC0105  &    2  & 0.097& 8.38 &  8.368& 0.088& 8.420& 0.185 \\
SMC0074  &    5  & 0.094& 8.04 &  7.957& 0.222& 7.916& 0.163 \\
SMC0068  &    2  & 0.101& 8.30 &  8.315& 0.032& 8.270& 0.018 \\
SMC0216  &    3  & 0.194& 8.22 &  8.199& 0.128& 8.211& 0.177 \\
SMC0159  &    2  & 0.084& 8.31 &  8.367& 0.119& 8.336& 0.122 \\
SMC0158  &    3  & 0.084& 8.00 &  8.038& 0.103& 7.972& 0.118 \\
SMC0149  &    4  & 0.079& 8.04 &  8.111& 0.053& 8.051& 0.064 \\
\enddata			
\tablenotetext{a}{Cluster identification according to Pietrzynski \& Udalski 
(1999).}
\tablenotetext{b}{Number of cluster Cepheids.}
\tablenotetext{c}{Mean cluster reddening based on individual cluster 
Cepheid reddenings (Pietrzynski \& Udalski 1999).  }
\tablenotetext{d}{Cluster age based on stellar isochrones (yr). The uncertainty 
on this estimate is of the order of $\sigma_{ISO}=0.075$ dex. } 
\tablenotetext{e}{Mean cluster age based on fundamental and first 
overtone PA relations (yr).   }
\tablenotetext{f}{Standard deviation.  } 
\tablenotetext{g}{Mean cluster age based on fundamental and first overtone 
PAC relations (yr).   }
\tablenotetext{h}{Standard deviation.}    
\end{deluxetable}

%7========================== Table X ================================
\clearpage 
\begin{deluxetable}{lcccccccc}
\normalsize
%\scriptsize 
\tablecolumns{9} 
\tablewidth{0pt} 
\tablecaption{Age estimates for two Galactic open clusters that host at 
least two Cepheids.\label{t2}}
%\scriptsize 
\tablehead{
\colhead{ID\tablenotemark{a}}&
\colhead{$\log P$\tablenotemark{b}}&
\colhead{$<B>$\tablenotemark{c}}&
\colhead{$<V>$\tablenotemark{c}}&
\colhead{$<I>$\tablenotemark{c}}&
\colhead{$E(B-V)$\tablenotemark{d}}&
\colhead{$\log t_{PA}$\tablenotemark{e}}&
\colhead{$\log t_{PAC}^{(B-V)}$\tablenotemark{f}}&
\colhead{$\log t_{PAC}^{(V-I)}$\tablenotemark{g}} \\ 
\colhead{(1)}&
\colhead{(2)}&
\colhead{(3)}&
\colhead{(4)}&
\colhead{(5)}&
\colhead{(6)}&
\colhead{(7)}&
\colhead{(8)}&
\colhead{(9)}}
\startdata
\multicolumn{9}{c}{NGC~7790\tablenotemark{h} }\\			
CEa~Cas  &  0.711 & 12.070 & 10.920 & \ldots & 0.562 & 7.834 & 7.788 & \ldots \\  
CEb~Cas  &  0.651 & 12.220 & 11.050 & \ldots & 0.548 & 7.874 & 7.827 & \ldots \\  
CF~Cas   &  0.688 & 12.335 & 11.136 & 9.754 & 0.531 & 7.849 & 7.824 & 7.838 \\  
\multicolumn{9}{c}{NGC~6067\tablenotemark{h} }\\			
QZ~Nor   &  0.578 & 9.774  & 8.866 & 7.893 & 0.249 & 7.923 & 7.872 & 7.865 \\ 
V340~Nor &  1.053 & 9.526  & 8.375 & 7.151 & 0.315 & 7.605 & 7.704 & 7.708 \\  
\enddata			
\tablenotetext{a}{Cluster variable identification.   
\hspace*{2.5mm} $^b$Logarithmic Period [days]. 
\hspace*{2.5mm} $^c$Mean B,V,I magnitudes (mag).
\hspace*{2.5mm} $^d$Reddening.   
\hspace*{2.5mm} $^e$Age estimate based on the PA relation (yr).   
\hspace*{2.5mm} $^f$Age estimate based on the PAC$_{(B-V)}$ relation.   
\hspace*{2.5mm} $^g$Age estimate based on the PAC$_{(V-I)}$ relation.   
\hspace*{2.5mm} $^h$Photometry and reddening estimates for cluster variables
from Berdnikov (2000) [NGC~7790], and Laney \& Stobie (1994) [NGC~6067].}    
\end{deluxetable}

\end{document}